\documentclass{article}
\usepackage{float}
\usepackage{bbm}
\usepackage{listings}
\usepackage{booktabs} 
\usepackage{geometry} 
\usepackage{graphicx} 
\usepackage{xcolor, soul,url}  
\usepackage{amssymb}
\usepackage{hyperref}
\usepackage{tabularx}
\usepackage{enumitem}

\usepackage[
    backend=biber,
    citestyle=numeric,
    bibstyle=vancouver,
    sorting=none,
    natbib=true,
    useprefix=true,
    maxcitenames=2,
    mincitenames=1,
    uniquelist=false,
maxbibnames=999,
minbibnames=999
]{biblatex}
 \usepackage{amsmath}
 \usepackage{amsthm}
 \usepackage{algorithm}
\usepackage{algpseudocode}
\usepackage{pdflscape}
\usepackage{multirow}

\usepackage{amssymb}

\makeatletter
\newcommand{\AlgText}[1]{%
  \Statex
  \noindent\hspace*{\ALG@tlm}%
  \parbox[t]{\dimexpr\linewidth-\ALG@tlm\relax}{%
    \setlength{\parindent}{0pt}%
    #1%
  }%
}
\makeatother
\newcommand\independent{\protect\mathpalette{\protect\independenT}{\perp}}
\def\independenT#1#2{\mathrel{\rlap{$#1#2$}\mkern2mu{#1#2}}}
 
\title{Explored Ideas MSE change FDR}
\usepackage{listings}
\usepackage{xcolor}
\definecolor{darkgrey}{rgb}{0.25,0.25,0.25} 
\definecolor{codegreen}{rgb}{0,0.6,0}
\definecolor{codegray}{rgb}{0.5,0.5,0.5}
\definecolor{codepurple}{rgb}{0.58,0,0.82}
\definecolor{backcolour}{rgb}{0.95,0.95,0.92}

\lstdefinestyle{Rstyle}{
    language=R,                         
    backgroundcolor=\color{backcolour},  
    commentstyle=\color{codegreen},       
    keywordstyle=\color{magenta},         
    numberstyle=\tiny\color{codegray},      
    stringstyle=\color{codepurple},       
    basicstyle=\footnotesize\ttfamily,    
    breakatwhitespace=false,              
    breaklines=true,                      
    captionpos=b,                         
    keepspaces=true,                      
    numbers=left,                         
    numbersep=5pt,                        
    showspaces=false,                     
    showstringspaces=false,               
    showtabs=false,                       
    tabsize=2                             
}

\usepackage{xcolor}
\usepackage{tikz}
\usepackage{subcaption}

\newif\ifqfive
\qfivefalse
\ifqfive
  \def\qlevel{0.05}
\else
  \def\qlevel{0.10}
\fi

\tikzset{
  venncircle/.style={line width=1.1pt},
  venncount/.style={font=\bfseries\small, inner sep=1pt},
  vennlabel/.style={font=\small, align=center}
}

\newcommand{\ThreeMethodVenn}[7]{%
\begin{tikzpicture}[scale=0.92, every node/.style={transform shape}]

  \fill[blue!55, fill opacity=0.20]
    (-1,0.55) circle (1.65);

  \fill[red!60, fill opacity=0.20]
    (1,0.55) circle (1.65);

  \fill[green!55!black, fill opacity=0.20]
    (0,-0.85) circle (1.65);

  \draw[blue!70!black, venncircle]
    (-1,0.55) circle (1.65);

  \draw[red!75!black, venncircle]
    (1,0.55) circle (1.65);

  \draw[green!45!black, venncircle]
    (0,-0.85) circle (1.65);

  \node[vennlabel, text=blue!70!black] at (-2.12,2.30)
    {Proposed method\\XGBoost};

  \node[vennlabel, text=red!75!black] at (2.12,2.30)
    {Proposed method\\Lasso};

  \node[vennlabel, text=green!40!black] at (0,-2.70)
    {Dai's method};

  \node[venncount] at (-1.58, 0.82) {#1}; 
  \node[venncount] at ( 1.58, 0.82) {#2}; 
  \node[venncount] at ( 0.00,-1.63) {#3}; 

  \node[venncount] at ( 0.00, 1.28) {#4}; 
  \node[venncount] at (-0.70,-0.43) {#5}; 
  \node[venncount] at ( 0.70,-0.43) {#6}; 

  \node[venncount] at ( 0.00, 0.16) {#7}; 

\end{tikzpicture}%
}
\newtheorem{proposition}{Proposition}
\begin{document}
\begin{refsection}

\title{\bfseries\sffamily{False Discovery Rate Control via Prediction-Based, Model-Agnostic Data Splitting: An Application to Gut Microbiome Data in Liver Function}} 
\author{
{\bfseries Margaux Delporte$^{1,2}$}
$\quad$
$\quad$
{\bfseries Yushu Shi$^{1^*}$}\\[2mm]
$\quad$
\small\em $^1$  Department of Population Health Sciences, Weill Cornell Medicine, New York, NY 10065, US\\
\small\em $^2$  Department of Public Health, University of Rhode Island, Kingston, RI, 02881, US\\
\small\em $^*$  Corresponding author: yus4011@med.cornell.edu}
\date{}
\maketitle

\noindent\textbf{Abstract:}
Many methods have been proposed to control the false discovery rate (FDR) in feature selection, yet challenges remain when the relationship between predictors and the outcome is complex. We propose a prediction-based data-splitting procedure that can leverage flexible predictive models while controlling the FDR. The data are divided into three parts: one for model training and two for constructing the mirror statistic. For each feature, we construct a negative control using the remaining features without incorporating outcome information and define a test statistic based on the resulting change in predictive performance. Under the null, such a perturbation is approximately equally likely to increase or decrease predictive performance, yielding a statistic symmetric about zero. In contrast, perturbing an informative feature tends to reduce predictive performance, resulting in a large positive statistic. Features are selected by thresholding these statistics, with false discoveries estimated from the corresponding negative tail to control the FDR.
We apply the proposed method to liver function microbiome data, where high-dimensional, correlated features and potentially complex feature--outcome relationships make reliable type I error control particularly important for reproducible scientific discovery. Through simulation studies and the real-data application, we demonstrate that the proposed framework provides robust FDR control and can be integrated with state-of-the-art machine-learning models to accommodate complex relationships between features and the outcome.
 

\section{Introduction}\label{s:intro}
Advances in biology have enabled studies that measure hundreds or thousands of genomic variables, such as gene expression levels, immune signatures, or microbiome features, to investigate associations with a response. 
In modern genomic studies, investigators often aim to identify a minimal set of features that is reliably associated with a response, while controlling the proportion of false discoveries among the selected findings at a prespecified threshold. 
A widely adopted measure of type I error in feature-association studies is the false discovery rate (FDR), defined as the expected false discovery proportion (FDP). 
 
	\[
		\mathrm{FDR} = \mathbb{E}[\mathrm{FDP}]=\mathbb{E}\left[\frac{V}{R\vee 1}\right],
		\]
where $V$ is the number of false positives and $R$ is the number of selected features.

\paragraph{An overview of FDR control methods.}

One popular class of FDR-controlling methods is based on feature-specific \(p\)-values, with the Benjamini--Hochberg (BH) procedure being one of the most widely used approaches \cite{Benjamini1995ControllingMultiple}. The original BH procedure was shown to provide finite-sample FDR control under independence of the \(p\)-values. \textcite{BenjYeku01} subsequently showed that the BH procedure remains valid under certain positive-dependence structures, and proposed the Benjamini--Yekutieli (BY) procedure, which provides FDR control under arbitrary dependence among the \(p\)-values. Other \(p\)-value-based approaches include Storey's \(q\)-value framework \cite{Storey2002ARates}, which estimates the proportion of true null hypotheses from the empirical distribution of the \(p\)-values to improve power, and the two-stage adaptive procedure of \textcite{Benjamini2006AdaptiveRate}.

A parallel line of research focuses on features rather than on \(p\)-values and targets conditional rather than purely marginal feature relevance. 
A representative method is the knockoff framework introduced by \textcite{Barber2015ControllingKnockoffs} and extended to Model-X knockoffs by \textcite{Candes2018PanningSelection}. 
For each original feature, an artificial ``knockoff'' feature is constructed to serve as a negative control while preserving appropriate dependence relationships with the original features. Feature-importance statistics are then constructed by contrasting each original feature with its knockoff counterpart, and a data-dependent threshold is used to select features while controlling the FDR. 

A practical challenge of (Model-\(X\)) knockoffs is that exact FDR control requires the joint distribution of the features to be known, or sufficiently accurately estimated, so that valid knockoffs can be generated. Several extensions have therefore been developed to construct knockoffs for more flexible feature distributions, including approaches based on deep generative models \cite{RomaSesi19}, generative adversarial networks (GANs) \cite{JordYoon19}, and autoencoders \cite{LiuZhen19,ZhuFan21}. Although these methods broaden the range of feature distributions for which knockoffs can be constructed, their validity and power depend on how accurately the learned generator approximates the required exchangeability structure, and their implementation can involve additional architectural, optimization, and hyperparameter choices. As a consequence, the second-order Gaussian Model-\(X\) knockoff procedure remains one of the most widely used implementations \cite{Candes2018PanningSelection}.

Data splitting has a long history as a general statistical technique. It has been widely used for model assessment through procedures such as cross-validation \cite{Stone1974Cross-ValidatoryPredictions}, and for separating model construction from subsequent statistical inference \cite{Moran1961DividingDilemma,Cox1975BiometrikaLevels,Wasserman2009High-dimensionalSelection}. 
More recently, data splitting has been adapted specifically for FDR control. \textcite{Dai2023FalseSplitting} proposed a data-splitting procedure that constructs a feature-specific mirror statistic from two independently estimated regression coefficients. In general, the mirror statistic can be written as a signed function of the magnitudes of the two coefficient estimates, with the sign determined by whether the two estimates agree in direction. Consequently, when the two coefficient estimates have the same sign, the mirror statistic is positive, whereas disagreement in their signs produces a negative statistic. The construction has two key properties: relevant features tend to produce large positive mirror statistics, whereas, under the null, the distribution of the mirror statistic is asymptotically symmetric about zero. The negative tail can therefore be used to estimate the number of false discoveries among features in the corresponding positive tail, yielding a simple thresholding rule for asymptotic FDR control. The authors further proposed multiple data splitting to reduce the instability associated with a single random split and improve selection power.

A limitation of the approach of \textcite{Dai2023FalseSplitting}, however, is that its construction relies on signed regression-coefficient estimates. Its original methodological development and theoretical FDR guarantees are established for linear regression and Gaussian graphical models. As a result, direct extension to predictive models that do not naturally produce signed feature-specific regression coefficients, such as random forests, gradient-boosted trees, or other general machine-learning algorithms, is nontrivial. This motivates the development of model-agnostic data-splitting procedures in which evidence for feature relevance can instead be quantified through changes in predictive performance.



\paragraph{Feature selection for microbiome data.}

We apply the proposed approach to a microbiome dataset comparing patients with cirrhosis and healthy individuals.
In microbiome association studies, identifying a small set of reproducibly associated taxa can improve interpretability and translational utility by prioritizing candidate taxa for subsequent mechanistic studies and potential microbiome-modulating interventions.
However, microbiome feature selection is particularly challenging because of its high dimensionality and compositionality. Standard multiple-testing procedures such as BH and conventional knockoff constructions do not, by themselves, account for the compositional constraint. Consequently, their direct application may lead to inappropriate inference. Several methods have therefore been developed specifically to accommodate the compositional structure of microbiome data.


An early approach addressing this issue was proposed by \textcite{SrinXue21}, who developed a two-step compositional knockoff filter based on a linear log-contrast model. Their method first performs compositional screening and subsequently applies fixed-\(X\) knockoffs to the reduced feature set, while incorporating the sum-to-zero constraint \(\mathbf{1}^{\top}\beta=0\) on the regression coefficients to account for compositionality. The same group later developed the MRKF method \cite{SrinXue23} for multivariate outcomes, which instead uses an additive log-ratio transformation and therefore requires the selection of a reference taxon. In a parallel line of research, \textcite{YuanJin24} modeled the log-transformed relative abundances under the sum-to-zero constraint \(\mathbf{1}^{\top}\beta=0\) to construct mirror statistics for data splitting.

These existing approaches nevertheless have their limitations. Methods based on an additive log-ratio transformation require the selection of a reference taxon and represent the remaining \(J-1\) taxa through their log-ratios relative to that reference. Consequently, both the analysis and the interpretation of the selected features will depend on the choice of reference. As noted by \textcite{MontPujo25} and \textcite{GreeMart21}, ``the choice of the reference frame is a crucial point, as all the subsequent analyses will depend on it.'' Sum-to-zero linear log-contrast formulations avoid the need to designate a particular reference taxon, but they generally are still developed within linear regression frameworks and therefore assume that the phenotype is related to the microbial composition through a linear combination of log-transformed relative abundances or log-ratios. In practice, microbiome--host relationships can be nonlinear and highly context-dependent. For example, \textit{Akkermansia muciniphila} can contribute to gut homeostasis through mucin-derived metabolites and cross-feeding of other commensal organisms, whereas under conditions such as low dietary fiber or a disrupted microbial community, excessive mucin degradation may compromise the intestinal mucus barrier and contribute to adverse effects \cite{TingEnge25}. Such context-dependent behavior motivates methods that can accommodate more flexible relationships between microbial features and host phenotypes.

Our work builds on the FDR control framework of \textcite{Dai2023FalseSplitting} and substantially extends it to accommodate complex feature--outcome relationships through flexible machine-learning models. The framework is model-agnostic, allowing the predictive model to be chosen based on the available sample size, feature dimensionality, and complexity of the underlying feature--outcome relationship. Rather than relying on regression coefficients, we generate negative controls and construct mirror statistics from the changes in predictive performance.
The remainder of the paper is organized as follows. Section \ref{sec:methods} describes the proposed methodology in detail. Section \ref{sec:simulation} presents simulation studies comparing our approach with existing methods across a range of settings, and Section \ref{sec:realDataApplicatiion} applies the proposed method to the motivating cirrhosis microbiome study. Section \ref{s:disc} concludes with a discussion of the main findings, limitations, and directions for future research.

\section{Method formulation}\label{sec:methods}

We begin by introducing the notation that will be used throughout the paper. 
Let $\mathbf{y} = (y_1, \ldots, y_n)^\top$ denote the observed response vector. For each $i \in \{1, \ldots, n\}$, the response $y_i$ is observed together with a $J$-dimensional covariate vector $(x_{i1}, \ldots, x_{iJ})$, and we assume that $y_1,\ldots,y_n$ are conditionally independent given their covariates.
For microbiome abundance data, the features are subject to the sum-to-one constraint, which means 
$\sum_{j=1}^J x_{ij}=1,$ $i=1,\dots,n.$ 
Our method differs from marginal tests in terms of the null hypothesis evaluated. A marginal test considers only \(H_{0j}: X_{j} \independent Y.\) 
In contrast, our method evaluates whether the population out-of-sample predictive performance remains unchanged after applying a perturbation initiated by feature \(j.\) Specifically, we test
\(
H_{0j}:\mathcal P_{\mathrm{orig}}=\mathcal P(j),
\) where \(\mathcal P_{\mathrm{orig}}\) and \(\mathcal P(j)\) denote the out-of-sample predictive performance under the original data distribution and after perturbation, respectively. For general non-compositional predictors, the perturbation may be applied to feature \(j\) while leaving the remaining features unchanged. For compositional microbiome data, however, changing a single component without adjusting the others would violate the unit-sum constraint. We therefore use a closure-preserving procedure, under which the remaining components are adjusted to retain a valid composition.
Depending on the outcome type, the performance metric may be \(R^2\), cross-entropy loss, or another appropriate measure. A deterioration in predictive performance after perturbation provides stronger evidence of feature relevance; the sign of the performance difference is defined according to whether larger or smaller values of the metric are preferable.

\subsection{Single split feature selection}
\label{sec:singleSplit}
First, we randomly partition the $n$ samples into three mutually exclusive index sets $U_1$, $U_2$, and $U_3$. We denote the responses and covariates in the corresponding subsets by
$(\boldsymbol{y}^{(1)}, \boldsymbol{X}^{(1)})$, $(\boldsymbol{y}^{(2)}, \boldsymbol{X}^{(2)})$, and $(\boldsymbol{y}^{(3)}, \boldsymbol{X}^{(3)})$,
where the superscripts $(1)$, $(2)$, and $(3)$ indicate the samples indexed by $U_1$, $U_2$, and $U_3$, respectively.
By construction, $U_1 \cap U_2=\varnothing,$ 
$U_2 \cap U_3=\varnothing,$ 
$U_1 \cap U_3=\varnothing,$ and $U_1 \cup U_2 \cup U_3 = [n]$. 
Feature importance is evaluated on the two test sets indexed by $U_1$ and $U_2$, while the subset indexed by $U_3$ is used exclusively for model fitting. This separation mitigates overfitting and provides a robust basis for computing feature importance on two independent test sets. 
The choice of prediction model is left to the user. In practice, we recommend first splitting the full dataset into training and testing sets, fitting several candidate models such as tree-based methods, nonparametric spline models, and deep neural networks, evaluating predictive performance, and choosing the highest-performing model for the subsequent feature selection.


When larger values of \(P\) indicate better predictive performance, as with \(R^2\), the perturbed performance \(P_j\) for a null feature is approximately equally likely to be above or below the original performance \(P\). In contrast, if $X_j$ is predictive of $Y$, replacing \(X_j\) with its negative control typically reduces predictive accuracy; consequently, $\mathcal P_j$ tends to be smaller than $\mathcal P$. 
If smaller values indicate better predictive performance, as with cross-entropy loss, the sign should be reversed accordingly.
For feature $j,$ we compute the difference in the model performance 
before and after perturbation of each test set $\beta_j^{(1)}=\mathcal P^{(1)}- \mathcal P_j^{(1)},$ 
$\beta_j^{(2)}=\mathcal P^{(2)}- \mathcal P_j^{(2)}.$

For continuous outcomes, the coefficient of determination can be used to quantify predictive performance, with $R^2 = 1-\sum_i(y_i-\widehat{y}_i)^2/\sum_i(y_i-\bar{y})^2$. For binary outcomes, cross-entropy loss can be used, defined as $L=-\frac{1}{n}\sum_{i=1}^n\left[y_i\log(\widehat{p}_i)+(1-y_i)\log(1-\widehat{p}_i)\right],$ where $\widehat p_i$ is the predicted probability of taking the value $1.$
These performance metrics are model-agnostic and can be used with a broad range of predictive models, including neural networks, random forests \cite{Ho95}, and XGBoost \cite{ChenGues16}. Similarly, for time-to-event outcomes, survival-specific prediction metrics, such as the integrated Brier score \cite{GrafSchm99} or Uno's C-index \cite{UnoCai11}, can be considered.

 Given $\boldsymbol{\beta}^{(1)}=\{\beta_{j}^{(1)}\}_{j=1}^J$ and $\boldsymbol{\beta}^{(2)}=\{\beta_{j}^{(2)}\}_{j=1}^J,$ we construct the mirror statistic:
 \begin{equation}\label{eq:mirror}
    M_j=\text{sign}(\beta_j^{(1)}\beta_j^{(2)})(\vert\beta_j^{(1)}\vert+\vert\beta_j^{(2)}\vert), \quad j=1,\dots,J .
\end{equation}

A “mirror statistic” or “symmetric statistic” should have the following two key properties \cite{Dai2023FalseSplitting}:
\begin{enumerate}
    \item 
Larger positive values provide stronger evidence that the feature is relevant.
  \item  For any null feature, the sampling distribution of the mirror statistic is symmetric about zero.
\end{enumerate}

Under the null, positive mirror statistics can occur by chance, but their frequency is estimated using the symmetric negative tail of the null distribution.
It is worth noting that with a two-way split, i.e., using one half of the data to fit the model and the other half to compute feature importance, the difference in coefficients of determination, $\beta_j = \mathcal P-\mathcal P_j$, also satisfies the two requirements above and can be used as a mirror statistic. However, our simulation trials suggest that a three-way split with a large training set yields better empirical performance.






\subsection{Construction of the negative controls for compositional microbial data}
\label{sec:constructNegative}
Our method relies on prediction difference using the original features versus constructed alternative features. 
A simple and computationally efficient approach for constructing alternative features is to permute the observed values of each feature. A limitation of permutation is that it disrupts the correlation structure between the permuted feature and the remaining covariates. 
Another approach is to construct a feature-specific negative control using the remaining features. 
Practically, this can be done by computing the leading \(K\) principal components of the remaining \(J-1\) features and then using these components in a regression model to predict the \(j\)th feature. The predicted values from the regression model are used to construct the negative control for feature \(j\).
For the rest of this subsection, we consider a procedure tailored to compositional microbiome data. 

\begin{enumerate}[leftmargin=*]

\item For subject \(i\), define
\(
\mathbf r_{i,-j}
=
\mathcal C(\mathbf x_{i,-j})
=
\frac{\mathbf x_{i,-j}}
{\sum_{\ell\neq j}x_{i\ell}}
=
\frac{\mathbf x_{i,-j}}
{1-x_{ij}},
\)
where \(\mathcal C(\cdot)\) denotes the closure operation and
\(
\sum_{k\neq j}r_{ik}^{(-j)}=1.
\)
Thus, \(\mathbf r_{i,-j}\) represents the relative allocation of abundance among the remaining \(J-1\) features. This representation separates the abundance of feature \(j\) from the internal composition of the remaining features. The construction assumes \(x_{ij}<1\), so that the normalized remainder is well defined.

\item We estimate the conditional distribution of \(X_{j}\) given the normalized remainder \(\mathbf r_{i,-j}\), without incorporating the outcome \(Y\). 
To map the normalized remainder $\mathbf r_{i,-j}$ from the simplex to unconstrained Euclidean coordinates, we use its isometric log-ratio (ILR) transformation.
\(
\mathbf u_{i,-j}
=
\operatorname{ILR}(\mathbf r_{i,-j})
\in\mathbb R^{J-2}.
\)
When $0$ is present in $\mathbf r_{i,-j},$ we add a very small number to the zero-entry before the ILR transformation. 
We first learn the loading matrix from the training set, and then compute the first \(K\) principal components of \(\mathbf u_{i,-j}\) from the training set. Let \(\mathbf c_{ij}\in\mathbb R^K\) denote the resulting principal-component scores. 
Conditional on \(\mathbf c_{ij}\) and, when applicable, additional nonmicrobial covariates \(\mathbf z_i\), we model the distribution of \(X_{ij}\) from the training data using a Beta model:
\[
X_{ij}\mid\mathbf c_{ij}, \mathbf z_i
\sim
\operatorname{Beta}
\left(
\mu_{ij}\phi_j,\,
(1-\mu_{ij})\phi_j
\right),
\]
where $\mu_{ij}\in(0,1)$ denotes the conditional mean and
$\phi_j>0$ is a feature-specific precision parameter. We model the
conditional mean through a logit link,
\(
\operatorname{logit}(\mu_{ij})
=
\alpha_j+\mathbf z_i^{\top}\boldsymbol{\delta}_j+
\mathbf c_{ij}^{\top}\boldsymbol{\gamma}_j.
\)
For each subject in the held-out evaluation sets, we 
apply the loading matrix and Beta regression coefficients learned from the training set to
generate a stochastic conditional draw
\(
\widetilde x_{ij}
\) from the fitted Beta regression model.

\item We reconstruct the complete perturbed composition while preserving the relative allocation among the remaining \(J-1\) features. We 
define
$\widetilde{x}^{(j)}_{ij}=\widetilde{x}_{ij},$
\(
\widetilde x_{ik}^{(j)}
=
(1-\widetilde x_{ij})r_{ik}^{(-j)}
=
x_{ik}
\frac{1-\widetilde x_{ij}}
{1-x_{ij}} 
\) \(k\neq j.\)

\end{enumerate}

\subsection{Multiple Data Splitting}
Single data splitting may suffer from reduced statistical power relative to methods that use the full dataset, as well as instability in feature selection across different random splits.
\textcite{Dai2023FalseSplitting} addressed these two concerns through 
multiple data splitting (MDS). 
In this subsection, we further extend MDS for prediction-based and model-agnostic data splitting.

Given a dataset $(\mathbf{X}, \mathbf{y})$, we independently generate $R$ random data partitions. We recommend using $R=10$ in practice. For each partition, the observations are divided into $B$ approximately equal-sized folds. We recommend setting $B=10$.
For each partition, $B/2=5$ rotations are performed, with $80\%$, $10\%$, and $10\%$ of the data allocated to the training and two evaluation sets, respectively. 
 If a different value of $B$ is used, the proportions allocated to the training and evaluation sets may need to be adjusted accordingly.
The evaluation folds are rotated such that each fold contributes to the mirror-statistic calculation once within each partition.
Each dataset therefore yields a total of \(D=10\times 5=50\) sets of mirror statistics. We recommend this repeated \(80\%/10\%/10\%\) scheme because our numerical experiments suggest that constructing a reliable predictive model using a larger training set can yield more stable and informative feature-importance measures.

For each individual data split, the selected features form a set $\widehat{S}^{(d)}$, $d=1,\dots,D.$ After completing all splits, we compute the empirical inclusion rate $\widehat{I}_j$ for each feature $j=1,\dots,J$, which quantifies the stability and importance of that feature across repeated splits. Motivated by the premise that features selected less frequently across repeated splits are less likely to represent true signals, we use features with low inclusion rates to approximate false discoveries.
Let $\widehat{S}^{c}=\{j:\widehat{I}_j\leq \widehat{I}_{(L)}\}$ denote the set of features excluded from the final selection, where $\widehat{I}_{(1)}\leq\cdots\leq\widehat{I}_{(J)}$ are the ordered inclusion rates. The criterion $\sum_{\ell=1}^{L}\widehat{I}_{(\ell)}\leq q$ is equivalent to $\frac{1}{D}\sum_{d=1}^{D}\frac{\left|\widehat{S}^{(d)}\cap\widehat{S}^{c}\right|}{\left|\widehat{S}^{(d)}\right|\vee 1}\leq q$. Thus, the average proportion of split-specific selections that fall into the final excluded set, which serves as a proxy for the false discovery proportion, is controlled at or below the target level $q$.
Empirically, MDS tends to yield an FDP below the nominal level while retaining good power. Full details are provided in Algorithm~\ref{alg:repeated_split}.


\begin{algorithm}[!t]
\caption{FDR control via prediction-based multiple data splitting}
\label{alg:repeated_split}
\begin{algorithmic}
\AlgText{\textbf{Require:} Data $(\mathbf X, \mathbf y)$,
nominal FDR level $q\in(0,1)$, number of folds $B=10$,
number of random data partitions $R=10,$ and number of principal components $K=5.$}
\For{$r=1,\dots,R$}
\AlgText{
Randomly partition the data into $B$ approximately equal-sized folds
$(\mathbf X^{(1)},\mathbf y^{(1)}),\allowbreak \dots,\allowbreak
(\mathbf X^{(B)},\mathbf y^{(B)})$.
}
\For{$b=1,\dots,B/2$}
\AlgText{
Use $(\mathbf X^{(b)},\mathbf y^{(b)})$ and
$(\mathbf X^{(b+B/2)},\mathbf y^{(b+B/2)})$ as evaluation sets and
the rest of the data as the training set. Build a predictive model
$\widehat F_{b,r}$ using only the training set.
}
\AlgText{
Using the fitted model $\widehat F_{b,r},$ compute the original prediction-performance
measures $\mathcal P_{b,r}^{(1)}$ and $\mathcal P_{b,r}^{(2)}$ for
$(\mathbf X^{(b)},\mathbf y^{(b)})$ and
$(\mathbf X^{(b+B/2)},\mathbf y^{(b+B/2)})$, respectively.
}
\For{$j=1,\dots,J$}
\AlgText{
For each evaluation dataset, compute the first $K$ principal component
scores based on the remaining $J-1$ features using the loading matrix
estimated from the training set.
}
\AlgText{
\scriptsize{\textbf{For compositional microbiome data:}
Apply the closure operation to the remaining features, perform an ILR transformation, and then compute the first \(K\) principal components.}
}
\AlgText{Construct a simple regression model $\widehat f_{j,b,r}$ from the remaining $J-1$ features in the training data with $K$ principal components as inputs for predicting the $j$th feature }
\AlgText{
Draw random predictions $\widetilde X_{j,b,r}^{(1)}$ and
$\widetilde X_{j,b,r}^{(2)}$ from $\widehat f_{j,b,r}$ as negative controls for the $j$th feature
using the first $K$ principal components of the remaining features in
the evaluation sets.}
\AlgText{
\scriptsize{\textbf{For compositional microbiome data:}
Apply a closure-preserving adjustment so that the perturbed observations
remain on the simplex.}
}
\AlgText{
Using the fitted predictive model $\widehat F_{b,r}$ to compute the post-perturbation
prediction-performance measures
$\widetilde{\mathcal P}_{j,b,r}^{(1)}$ and
$\widetilde{\mathcal P}_{j,b,r}^{(2)}$.
}
\AlgText{
When larger values of $\mathcal P$ indicate better predictive
performance,
define the prediction-performance changes as
\(
\beta_{j,b,r}^{(k)}=
\mathcal P_{b,r}^{(k)}-
\widetilde{\mathcal P}_{j,b,r}^{(k)},
\) $k\in\{1,2\}.$
Reverse the difference for smaller-is-better metrics.
}
\AlgText{
Compute the mirror statistic
\(
M_{j,b,r}\) using Equation~(\ref{eq:mirror}).
}
\EndFor
\AlgText{
Set the split-specific threshold
\(
\tau_{q,b,r}=
\min\left\{
t>0:
\frac{
\#\{j:M_{j,b,r}<-t\}
}{
\#\{j:M_{j,b,r}>t\}\vee1
}
\le q
\right\}.
\)
}
\AlgText{
Define the split-specific selected set
\(
\widehat S_{b,r}=
\{j:M_{j,b,r}>\tau_{q,b,r}\}.
\)
}
\EndFor
\EndFor
\For{$j=1,\dots,J$}
\AlgText{
Compute the inclusion rate across splits:
\(
\widehat I_j=
\frac{2}{BR}
\sum_{r=1}^{R}
\sum_{b=1}^{B/2}
\frac{
I\{j\in\widehat S_{b,r}\}
}{
|\widehat S_{b,r}|\vee1
}.
\)
}
\EndFor
\AlgText{
Sort the inclusion rates:
\(
0\le
\widehat I_{(1)}
\le
\widehat I_{(2)}
\le
\cdots
\le
\widehat I_{(J)}.
\)
}
\AlgText{
Find the largest $L\in\{1,\dots,J\}$ such that
\(
\sum_{\ell=1}^{L}
\widehat I_{(\ell)}
\le q.
\)
}
\AlgText{
Set the inclusion-rate threshold
\(
t_q=\widehat I_{(L)}.
\)
}
\AlgText{
\textbf{Return}
\(
\widehat S=
\left\{
j:\widehat I_j>t_q
\right\}.
\)
}
\end{algorithmic}
\end{algorithm}

\section{Simulation study}
\label{s:sim}

\label{sec:simulation}
We evaluated the proposed procedure using a full factorial design that varies dimensionality, outcome type, the functional form of the association, and signal sparsity. Specifically, we consider low-dimensional settings with \(J=100\) and high-dimensional settings with \(J=1000\). For both low and high-dimensional settings, the number of observations $n=1000.$ For each dimensional setting, we considered continuous and binary outcomes, linear and nonlinear associations, and two levels of relevant features. This results in \(2\times2\times2\times2=16\) core simulation scenarios. No external covariates were generated, so the outcome depends only on the simulated microbial composition. Each scenario is independently replicated \(200\) times.
We examined the performance at nominal FDR levels
\(
q\in\{0.05,0.10\},
\)
and computed the median and the $2.5$th and $97.5$th percentiles of FDP and empirical power across $200$ replicates for each simulation setting.

\subsection{Simulation setup}
\label{sec:simulationSetup}
\paragraph{Feature abundance generation.}
To evaluate feature selection performance under microbiome-like compositional structures, we generated microbial abundance profiles from a Dirichlet distribution. Specifically, for each subject $i$, the relative abundance vector $\mathbf{x}_i=(x_{i1},\ldots,x_{iJ})$ was generated as $\mathbf{x}_i\sim \mathrm{Dirichlet}(\alpha_1,\ldots,\alpha_J)$, where $\alpha_j=0.5$ for all features. This construction ensures that each observation satisfies the compositional constraint $\sum_{j=1}^{J}x_{ij}=1$.
\paragraph{Low-dimensional setting $J=100.$}
The number of true features was set to $6$ and $10.$ For continuous outcomes, the outcome noise was calibrated to achieve an oracle coefficient of determination of $R^2=0.20.$
\paragraph{High-dimensional setting $J=1000.$}
The number of true features was set to 10 and 20. For continuous outcomes, the outcome noise was calibrated to achieve an oracle coefficient of determination of $R^2=0.30$ and $R^2=0.60$, respectively, allowing us to evaluate feature selection performance under different signal-to-noise regimes while maintaining approximate equal contribution among true features.

\paragraph{Linear and nonlinear continuous outcome generation.}
For linear scenarios, we generated outcomes according to a log-contrast model to respect the compositional nature of microbiome abundance data. Specifically, for a prespecified set of signal pairs $\mathcal{S}$, we defined the linear predictor as $\eta_i=\sum_{h\in\mathcal{S}}\{\log(x_{i,2h-1})-\log(x_{i,2h})\}.$ 
For nonlinear scenarios, we first constructed pairwise log-ratio balances $B_{ih}=\{\log(x_{i,2h-1})-\log(x_{i,2h})\}/\sqrt{2}.$ To induce a non-monotone relationship, each selected balance entered the outcome through its squared value. We standardized each squared balance as $Q_{ih}^{*}=\{B_{ih}^{2}-   \frac{1}{n}\sum_{i=1}^{n}B_{ih}^2,\}/\operatorname{SD}(B_h^{2})$ and defined the latent signal as $\eta_i=\sum_{h\in\mathcal S}Q_{ih}^{*}$. Thus, each signal balance contributed on a comparable scale, while the dependence on $B_{ih}^{2}$ generated a U-shaped feature--response relationship that could not be represented by an additive linear model. Because the signal was defined through log-ratios, proportional rescaling of both components of a signal balance left $\eta_i$ unchanged, thereby preserving the interpretation of null features under the composition-preserving perturbation. The response was generated as $y_i=\eta_i+\epsilon_i$, where $\epsilon_i\sim N(0,\sigma^{2})$ and $\sigma^{2}=\operatorname{Var}(\eta)(1-R^{2})/R^{2}$ was chosen to achieve the prespecified oracle coefficient of determination. 

\paragraph{Binary outcomes generation.}
For the binary-outcome simulations, we first generated a continuous latent response using the same underlying signal structures as in the continuous-outcome simulations. The observed binary outcome was subsequently obtained by dichotomizing the continuous latent response at its sample median. This construction yielded approximately balanced binary classes. For the $J=100$ simulations, we set the target latent oracle $R^2$ to $0.30$ in both settings. For the $J=1000$ simulations, we set the target latent oracle $R^2$ to $0.45$ and $0.90$ when the number of true features was $10$ and $20$, respectively, to accommodate the greater dimensionality of the feature space.

\paragraph{Benchmark methods}
We implemented \textcite{BenjYeku01}'s method, knockoff, knockoff+, and \textcite{Dai2023FalseSplitting}'s multiple data splitting approach as benchmark methods. 
For \textcite{BenjYeku01}'s method, we used marginal p-values from linear regression and t-test for continuous and binary outcomes, respectively.
For knockoff and knockoff+ methods, we used 10-fold cross-validated Lasso\cite{Tibs96} with Lasso coefficient difference knockoff statistics.
For the method of \textcite{Dai2023FalseSplitting}, the data were randomly divided into two equal-sized subsets in each split, and we conducted $50$ splits for each replicate. On the first subset, we fitted a Lasso regression model with the penalty parameter selected using 10-fold cross-validation. Features with nonzero Lasso coefficients were retained as the screened set. On the second subset, ordinary least squares regression was fitted using only the screened features, and coefficients for nonscreened features were set to zero. Following the implementation in the \textcite{Dai2023FalseSplitting}’s GitHub repository, the mirror statistic for feature $j$ was defined as $W_j=\mathrm{sign}(\beta^{(1)}_j\beta^{(2)}_j)(|\beta^{(1)}_j|+|\beta^{(2)}_j|)$, where $\beta^{(1)}_j$ and $\beta^{(2)}_j$ denote the Lasso and OLS estimates (unpenalized logistic regression for binary outcomes) from the two independent data splits, respectively. 

\paragraph{Details on model parameters.}
For the proposed data-splitting procedure, we used XGBoost as the predictive model with final results aggregated from $50$ splits. Each XGBoost model consisted of 350 sequential boosting trees and was fitted using the squared-error objective (\texttt{reg:squarederror}) for continuous outcomes and the binary logistic classification objective (\texttt{binary:logistic}) for binary outcomes. Individual trees were restricted to a maximum depth of 2, and a learning rate of $\eta=0.04$ was used to gradually update the ensemble. To reduce overfitting, each tree was fitted using 85\% of the available training observations (\texttt{subsample}=0.85) and 85\% of the candidate predictors (\texttt{colsample\_bytree}=0.85). We specified a minimum child weight of 2, an $L_2$ regularization parameter of 2, and no additional $L_1$ regularization. 

\subsection{Simulation performance}
\label{sec:simulationPerformance}
\begin{figure}[H]
    \centering
    \includegraphics[width=\linewidth]{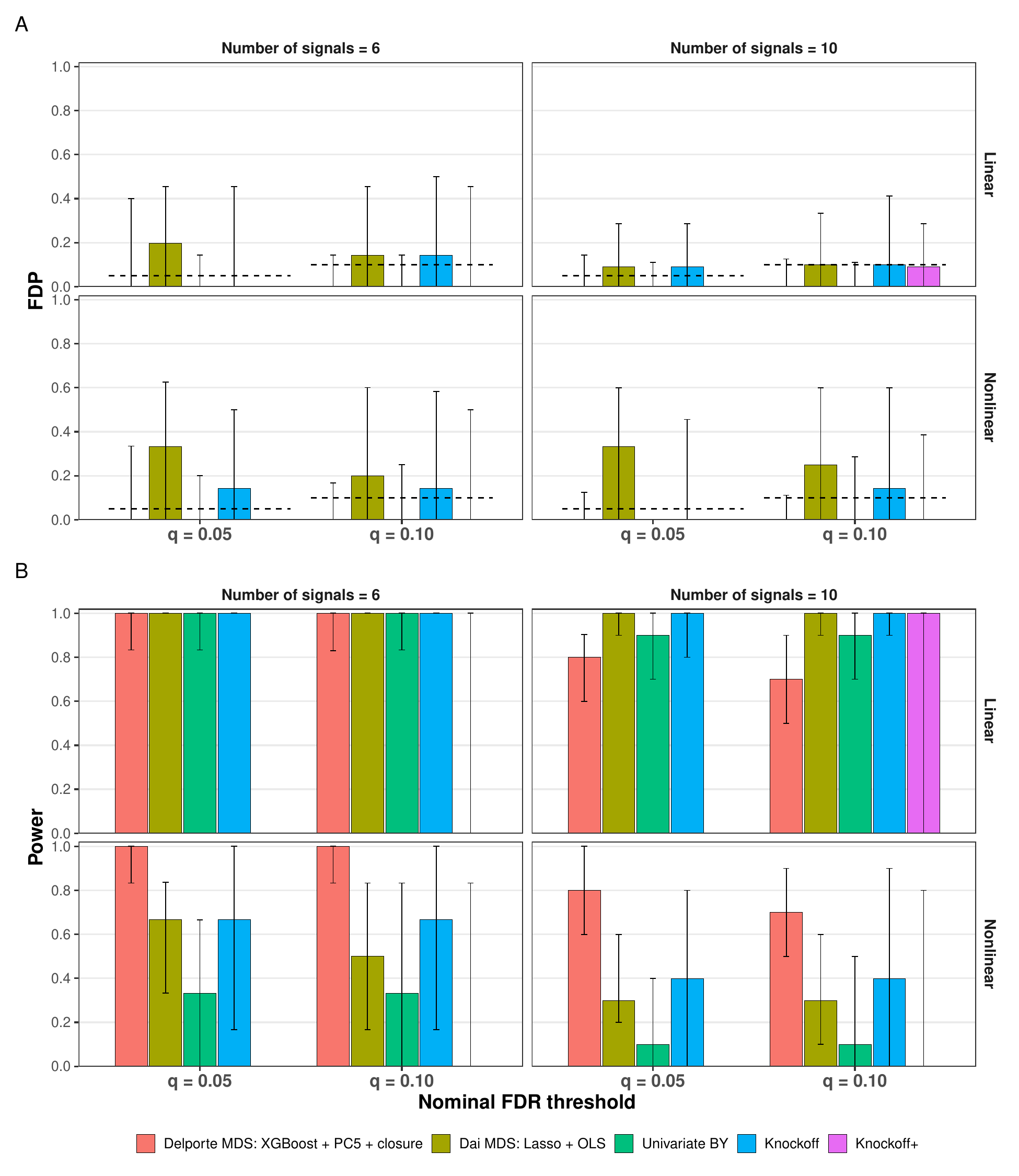}
    \caption{
    Median FDP (panel A) and median power (panel B) across 200 continuous-outcome simulation replicates with \(n=1000\) and \(J=100\). Columns correspond to 6 and 10 true signals, and rows within each panel correspond to linear and nonlinear signal–outcome relationships. Bars show medians, and error bars show the $2.5$th and $97.5$th percentiles. Dashed horizontal lines in panel A indicate the corresponding nominal FDR level \(q\). 
    }
    \label{fig:continuous100FDRPower}
\end{figure}

\begin{figure}[H]
    \centering
    \includegraphics[width=\linewidth]{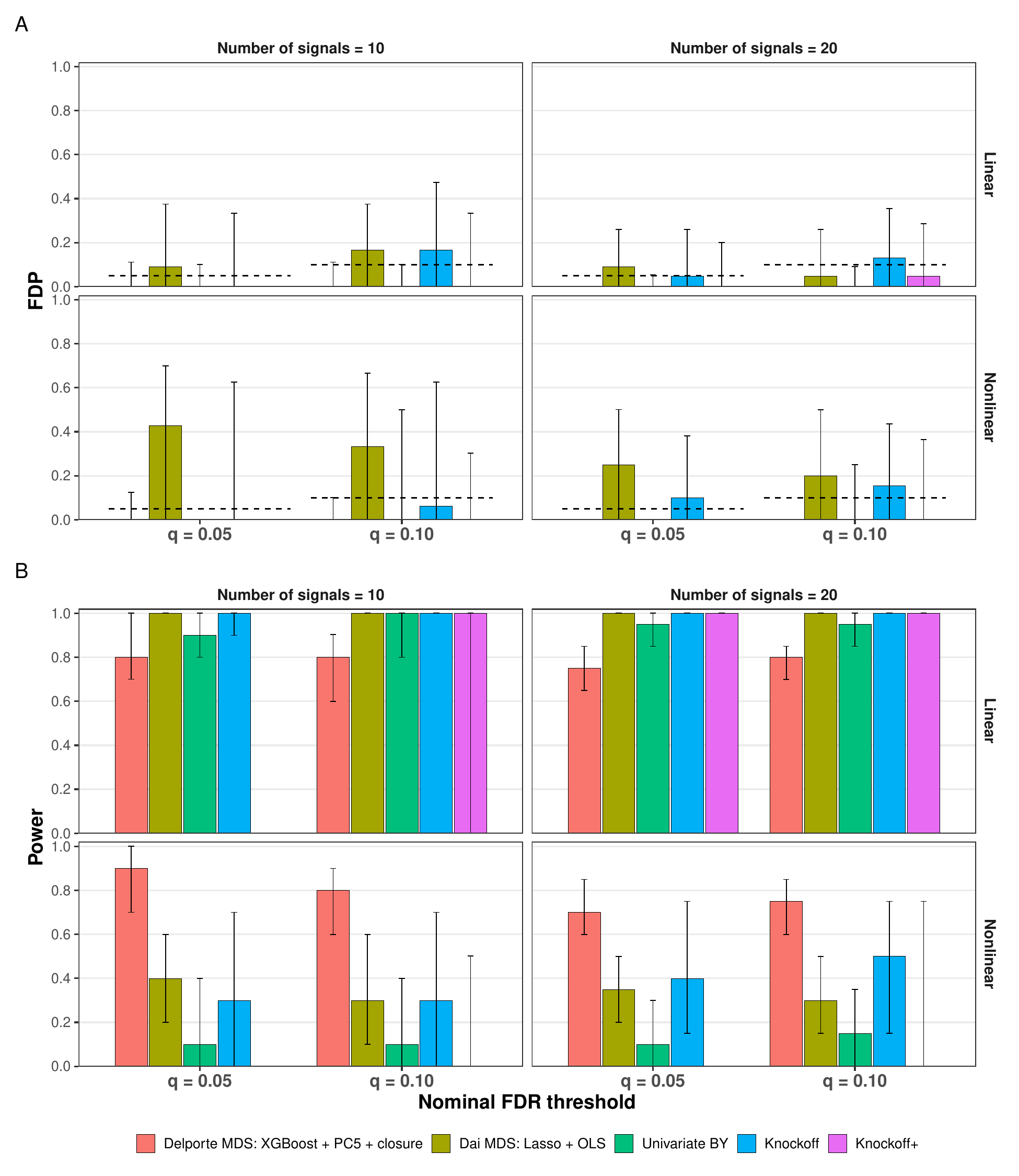}
    \caption{    Median FDP (panel A) and median power (panel B) across 200 continuous-outcome simulation replicates with \(n=1000\) and \(J=1000\). Columns correspond to 10 and 20 true signals, and rows within each panel correspond to linear and nonlinear signal–outcome relationships. Bars show medians, and error bars show the $2.5$th and $97.5$th percentiles. Dashed horizontal lines in panel A indicate the corresponding nominal FDR level \(q\).}
    \label{fig:continuous1000FDRPower}
\end{figure}

\paragraph{Continuous outcome simulation performance.}
As shown in Figure \ref{fig:continuous100FDRPower} and \ref{fig:continuous1000FDRPower}, the proposed method overall achieved a favorable balance between FDR control and power, particularly when the underlying association was nonlinear. In the linear settings, our method generally had moderate-to-high power, although its power was sometimes lower than that of \textcite{Dai2023FalseSplitting}'s, the knockoff method, and the univariate BY procedure. For example, when $J=1000$, the median power of our method was approximately $0.75$--$0.80$ under linear outcome models, whereas the method of \textcite{Dai2023FalseSplitting} and the standard knockoff procedure frequently achieved power close to $1$. 
Importantly, however, our method generally provided substantially stronger FDR control. For $J=1000$, its median FDPs ranged from essentially zero to approximately $0.01$ across the linear settings, well below both nominal FDR levels of $0.05$ and $0.10$. In contrast, the method of \textcite{Dai2023FalseSplitting} and the standard knockoff procedure frequently exceeded the nominal FDR level despite their high power.

The advantage of our method became substantially more pronounced under nonlinear outcome models. Its power was remarkably stable when moving from the linear to the nonlinear settings. For $J=100$, the median power of our method was identical between the corresponding linear and nonlinear scenarios: it was $1.00$ with six active features and $0.70$--$0.80$ with ten active features. A similar pattern was observed for $J=1000$, where median power was approximately $0.75$--$0.80$ under the linear models and $0.70$--$0.90$ under the nonlinear models. This relative insensitivity to the true underlying feature--outcome relationship is consistent with the use of XGBoost as the underlying predictive learner, which performs similarly when the underlying relationship is linear or nonlinear.

In contrast, the performance of several competing procedures deteriorated considerably when the outcome model became nonlinear. For $J=1000$, \textcite{Dai2023FalseSplitting}'s method had essentially perfect power under the linear models, but its median power decreased to only $0.30$--$0.40$ under the nonlinear models. The standard knockoff procedure showed a similar reduction, from approximately $1.00$ power in the linear settings to $0.30$--$0.50$ in the nonlinear settings. The univariate BY procedure was particularly sensitive to nonlinearity: although it achieved median power of approximately $0.90$--$1.00$ in the linear settings, its median power was only $0.10$--$0.15$ for $J=1000$ under nonlinear alternatives. Therefore, although these competing methods can be highly efficient when the underlying signal is well represented by their working models, their power can be substantially reduced under model misspecification. The corresponding loss of power for our method was much smaller.

This robustness was obtained while maintaining strong error control. In the nonlinear $J=1000$ settings, the FDP of our method was at most approximately $0.011$, whereas that of \textcite{Dai2023FalseSplitting}'s method ranged from approximately $0.20$ to $0.42$ and that of the standard knockoff procedure ranged from approximately $0.11$ to $0.17$. Thus, the greater power of our method under nonlinear alternatives did not result from selecting a larger number of false-positive features. The univariate BY and knockoff+ procedures also tended to control the FDR conservatively, but this was accompanied by substantial losses of power. In particular, knockoff+ frequently selected no variables in the sparse or nonlinear settings, resulting in median power equal to zero. These results illustrate the usual trade-off between stringent error control and discovery power, with our method providing a more useful compromise, especially when the functional form of the association is not known in advance.

For the $J=100$ continuous-outcome simulations, the apparent reduction in power in the setting with more true signals should not be interpreted as evidence that the proposed method becomes less effective as the number of true signals increases. The simulation design simultaneously increased the number of true signals from six to ten while keeping the total outcome variance explained by the signal fixed at $R^2=0.20$. Consequently, the average signal attributable to each individual active feature became weaker: heuristically, the available explained variation was distributed over ten rather than six features. The resulting decrease in feature-specific signal strength makes individual variables more difficult to detect. The observed reduction in power therefore reflects a more challenging signal configuration rather than an adverse effect of increasing sample size.

Taken together, the simulations suggest that the principal advantage of our method is not necessarily maximal power under an idealized
linear model. Rather, its strength lies in combining reliable FDR control with relatively stable power across different forms of the outcome model. \textcite{Dai2023FalseSplitting}'s method and the standard knockoff method can achieve greater power when the association is linear, but their advantages diminish substantially under nonlinear alternatives and are accompanied by less satisfactory FDR control in several scenarios. The proposed method therefore appears particularly attractive when the true exposure--outcome relationship is complex or unknown, which is likely to be the more realistic situation in high-dimensional applications.

An interesting observation is that the number of selected features at a more stringent FDR threshold ($q=0.05$) can occasionally be larger than that obtained at a less stringent threshold ($q=0.10$) in the multiple data splitting framework. This phenomenon does not indicate a violation of FDR control, but rather results from the stochastic nature of the aggregation and thresholding procedure. A detailed explanation is offered in Supplemental Section \ref{sec:numericalBuffer}.

\begin{figure}[H]
    \centering
    \includegraphics[width=\linewidth]{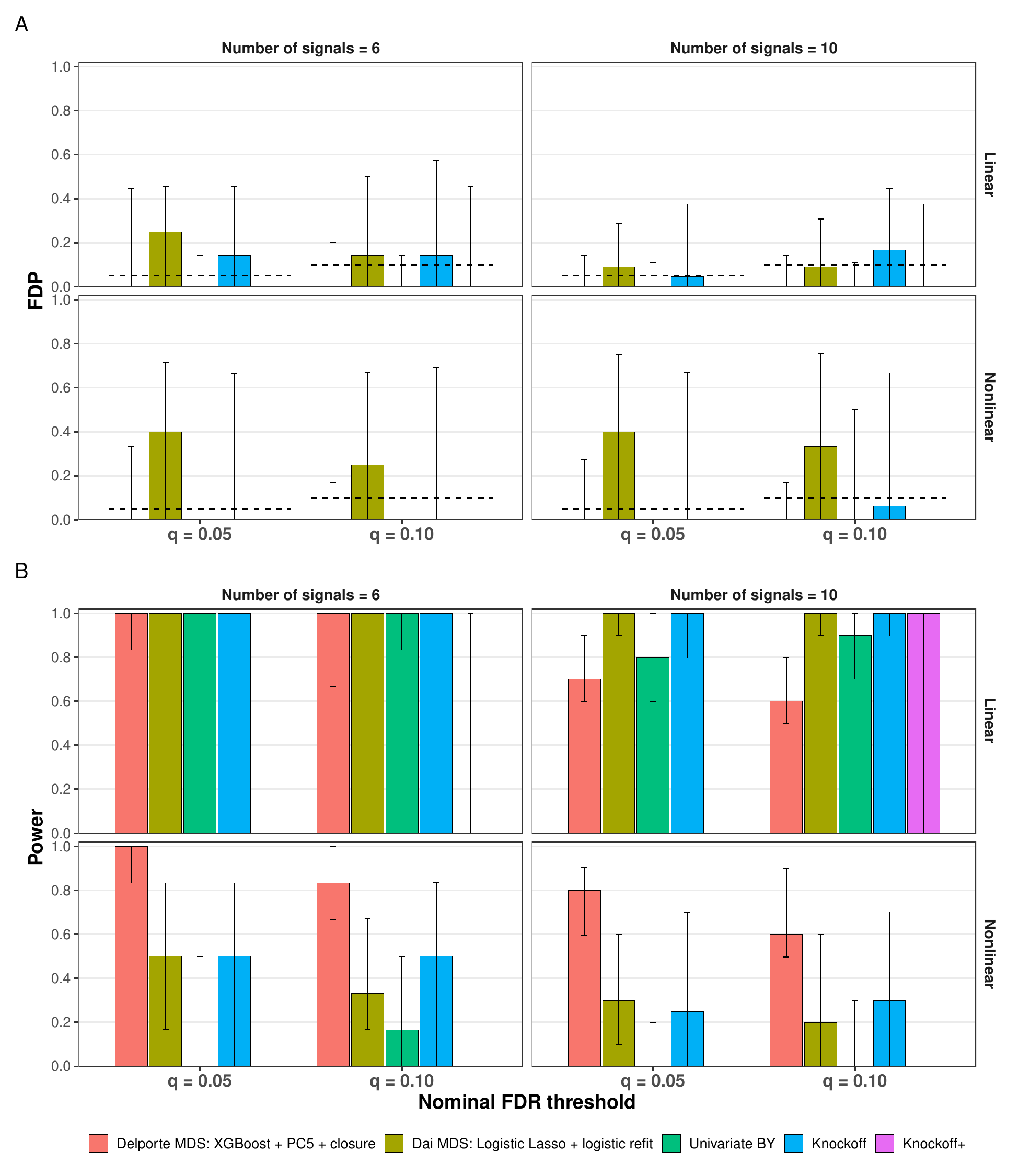}
    \caption{Median FDP (panel A) and median power (panel B) across 200 binary-outcome simulation replicates with \(n=1000\) and \(J=100\). Columns correspond to 6 and 10 true signals, and rows within each panel correspond to linear and nonlinear signal–outcome relationships. Bars show medians, and error bars show the $2.5$th and $97.5$th percentiles. Dashed horizontal lines in panel A indicate the corresponding nominal FDR level \(q\). }
\label{fig:binary100FDRPower}
\end{figure}

\paragraph{Binary outcome simulation performance.}
For binary outcomes, the proposed method demonstrated a favorable balance between false discovery rate (FDR) control and statistical power, particularly when the feature--outcome latent relationship was nonlinear. Under linear latent relationships, several existing methods achieved high power, as expected given that their underlying modeling assumptions were well aligned with the data-generating mechanism. As shown in Figure \ref{fig:binary100FDRPower}, when $J=100$, Dai's MDS, univariate BY, and the standard knockoff procedure frequently achieved power close to one. However, the increased power of Dai's MDS and the standard knockoff procedure was often accompanied by substantial FDR inflation. In contrast, the proposed method generally maintained a considerably lower FDR, although some finite-sample inflation was observed at the more stringent target $q=0.05$ in the sparsest settings. Univariate BY also maintained low FDR under the linear scenarios and, because the true latent relationship was linear, could achieve higher power than the proposed method.

The advantage of the proposed method became substantially more pronounced when the feature--outcome latent relationship was nonlinear. With $J=100$ and six true signals, the proposed method achieved powers of $0.979$ and $0.903$ at $q=0.05$ and $q=0.10$, respectively. In comparison, the corresponding powers were $0.466$ and $0.398$ for Dai's MDS, $0.100$ and $0.148$ for univariate BY, and $0.438$ and $0.443$ for the standard knockoff procedure. A similar pattern was observed with ten true signals: the proposed method achieved powers of $0.763$ and $0.647$, whereas the maximum power among Dai's MDS, univariate BY, and the standard knockoff procedure was only $0.290$ at $q=0.05$ and $0.301$ at $q=0.10$. Moreover, Dai's MDS and the standard knockoff procedure exhibited substantial FDR inflation in these nonlinear settings, whereas the proposed method produced much smaller FDRs.

\begin{figure}[H]
    \centering
    \includegraphics[width=\linewidth]{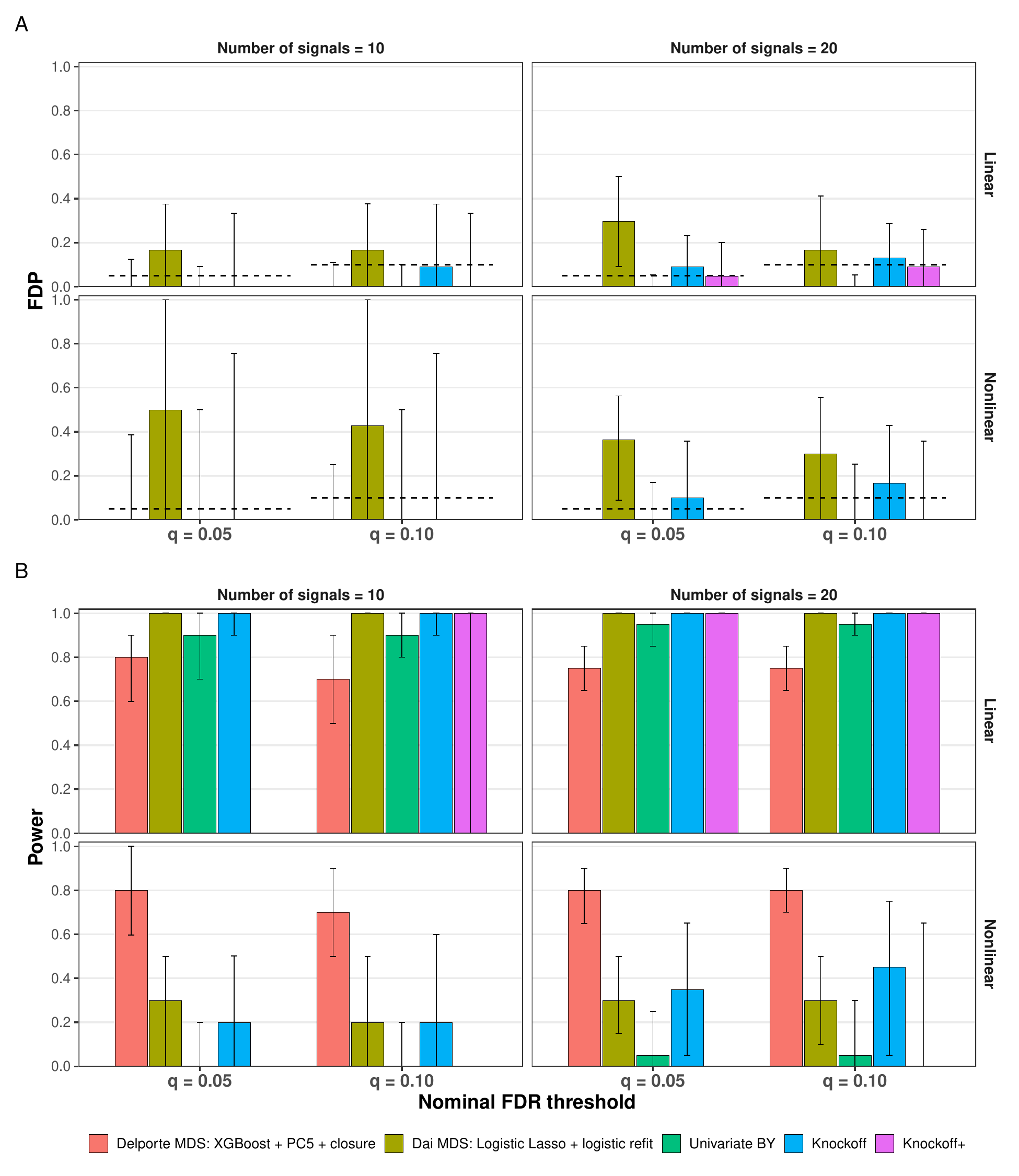}
    \caption{Median FDP (panel A) and median power (panel B) across 200 binary-outcome simulation replicates with \(n=1000\) and \(J=1000\). Columns correspond to 10 and 20 true signals, and rows within each panel correspond to linear and nonlinear signal–outcome relationships. Bars show medians, and error bars show the $2.5$th and $97.5$th percentiles. Dashed horizontal lines in panel A indicate the corresponding nominal FDR level \(q\). }
    \label{fig:binary1000FDRPower}
\end{figure}

The advantage under nonlinear latent relationships persisted and became particularly clear in the higher-dimensional setting with $J=1000$. As shown in Figure~\ref{fig:binary1000FDRPower}, with ten true signals, the proposed method achieved powers of $0.785$ and $0.679$ at $q=0.05$ and $q=0.10$, respectively, compared with $0.278$ and $0.234$ for Dai's MDS, $0.022$ and $0.032$ for univariate BY, and $0.209$ and $0.214$ for the standard knockoff procedure. With twenty true signals, the proposed method achieved powers of $0.783$ and $0.794$, while the corresponding powers were only $0.328$ and $0.301$ for Dai's MDS, $0.065$ and $0.094$ for univariate BY, and $0.357$ and $0.421$ for the standard knockoff procedure. Notably, in the latter setting the median FDP of the proposed method was zero at both nominal FDR levels, whereas Dai's MDS and the standard knockoff procedure substantially exceeded their nominal targets.

Knockoff+ was generally more conservative than the standard knockoff procedure but often suffered a substantial loss of power, particularly in sparse settings. For example, with ten true signals and $J=1000$, Knockoff+ selected no features under the linear setting at $q=0.05$, and its power was essentially zero under the corresponding nonlinear setting. Taken together, these results illustrate the principal benefit of the proposed model-agnostic framework: while methods built around linear or marginal associations can perform very well when their assumptions match the underlying relationship, their performance can deteriorate markedly under nonlinear alternatives. The proposed approach is substantially more robust to such departures and retains competitive power while generally providing considerably better FDR control than the less conservative multivariable competitors.

\paragraph{Computational time.}
We recorded the computational time on a MacBook Pro equipped with an Apple M2 Max chip for a simulation with a continuous outcome, $J=100$, and $n=1000$. The computational time required for each step of the proposed method, together with the computational time of the competing methods, are reported in Table~\ref{tab:computation_time}. For fixed model complexity and tuning parameters, the computational cost of both XGBoost and Lasso scales approximately as $\mathcal{O}(nJ)$, providing researchers with a practical basis for anticipating how computational time may increase with sample size and dimensionality. Because XGBoost is more computationally intensive than Lasso, the XGBoost-based implementation of our method requires a longer running time than several of the competing approaches. However, this additional computational cost is accompanied by stable FDR control and competitive power across a broad range of simulation settings.

\section{Real data application}
\label{sec:realDataApplicatiion}
We applied the proposed method to the dataset of \textcite{Qin2014AlterationsCirrhosis}, which contains stool samples from 123 cirrhosis patients and 114 healthy individuals. The metagenomic profiles were obtained from the \texttt{curatedMetagenomicData} R package \cite{Pasolli2017}. 
We considered two outcomes: serum albumin level, a key biomarker for which lower levels generally reflect worse liver function \cite{Jagdish2021AlbuminDrug}, as a 
continuous outcome and cirrhosis status as a binary outcome. 
To reduce noise from rare taxa, microbial features were filtered using established prevalence and abundance thresholds (present in $\geq10\%$ of samples, relative abundance $\geq1 \times 10^{-4}$, and exceeding this abundance in $\geq5\%$ of samples). After filtering, the dataset contained $204$ features. The code of the case study is available on \url{https://github.com/MargauxDelporte/FDR_Datasplitting}. 
Table~\ref{tab:qin_proposed_features} lists the microbial features identified 
by the proposed method.

For the Lasso-based implementation, we fitted a five-fold cross-validated Lasso model. For the XGBoost-based implementation, each model consisted of 300 trees. The maximum tree depth was set to $2,$ the learning rate to $0.05,$ the minimum child weight to $5,$ and both the row and feature subsampling proportions to $0.8.$ For albumin, we used a squared-error regression objective, whereas for cirrhosis, we used a binary logistic objective with cross-entropy loss.
For the continuous albumin outcome, the mean held-out $R^2$ values were $ 0.399$ for Lasso and $0.415$ for XGBoost, respectively. For the binary cirrhosis outcome, the corresponding mean held-out cross-entropy losses were $0.325$ for Lasso and $0.311$ for XGBoost.
These results indicate that, despite being more complex and potentially more challenging to fit with a relatively small sample size, XGBoost achieved predictive performance comparable to, and slightly better than, that of the linear Lasso model.

In the current analysis, all features selected at 
$q=0.10$ were also selected at $q=0.05$ for both predictive models and both 
outcomes. We therefore distinguish between features identified only at 
$q=0.05$ and those consistently identified at both nominal FDR levels.

\textit{Hungatella hathewayi} and \textit{Veillonella infantium} were selected for both outcomes by both learners at \(q=0.05\) and \(q=0.10\). \textit{Veillonella atypica} was selected for albumin by both learners, whereas \textit{Veillonella parvula} was selected for cirrhosis. The genus \textit{Veillonella} appeared repeatedly among the selected features for both albumin level and cirrhosis status.

Several microbial features identified by the proposed method have been 
previously implicated in liver cirrhosis. In particular, members of the genus \textit{Veillonella} \cite{JinCeni26}, including \textit{V. parvula} \cite{SungLin19}, have been repeatedly associated with cirrhosis and hepatic encephalopathy, potentially reflecting increased oral--gut microbial transmission during disease progression.
In addition, \textit{Hungatella} and 
\textit{Ruthenibacterium lactatiformans} have been reported in a recent study \cite{SharGupt24}
investigating cirrhosis complications. These multiple independent lines of evidence provide biological support for the ability of the proposed model-agnostic feature-selection approach to recover microbiome signatures consistent with findings from previous cirrhosis research.

\begin{figure}[!htbp]
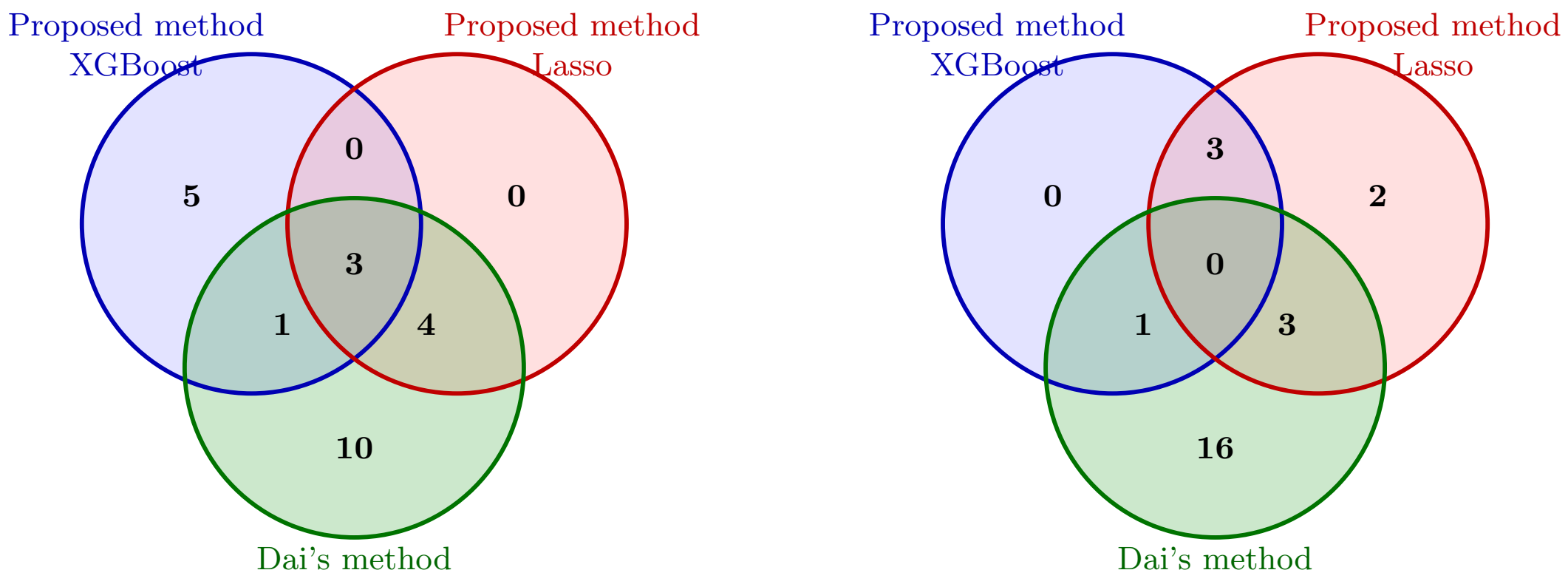

\centering

\begin{minipage}[t]{0.49\textwidth}
\centering

\ifqfive

  %

  \ThreeMethodVenn{7}{1}{11}{0}{3}{4}{4}

\else

  %

  \ThreeMethodVenn{5}{0}{10}{0}{1}{4}{3}

\fi

\subcaption{Albumin}
\label{fig:venn_albumin}
\end{minipage}
\hfill
\begin{minipage}[t]{0.49\textwidth}
\centering

\ifqfive

  %

  \ThreeMethodVenn{2}{3}{14}{3}{0}{4}{2}

\else

  %

  \ThreeMethodVenn{0}{2}{16}{3}{1}{3}{0}

\fi

\subcaption{Cirrhosis}
\label{fig:venn_cirrhosis}
\end{minipage}

\caption{
Overlap of microbial features selected by the proposed method with XGBoost,
the proposed method with Lasso, and Dai's method in the \textcite{Qin2014AlterationsCirrhosis} analysis
at $q=\qlevel$. Numbers denote the numbers of features in mutually exclusive
regions of the Venn diagrams.
}
\label{fig:qin_venn}

\end{figure}

We summarized the numbers of features selected by our method using Lasso and XGBoost, together with those selected by the competing methods, in Table \ref{tab:qin_selection_number}. The overall pattern is consistent with our simulation results: our method selects a moderate number of features, whereas the method of \textcite{Dai2023FalseSplitting} with lasso tends to select a larger number of features, although it also exhibits inflated Type I error in several simulation settings. In contrast, the knockoff methods with lasso select fewer features. We further compared the features selected by our method using Lasso and XGBoost with those selected by \textcite{Dai2023FalseSplitting} using Venn diagrams at $q=0.10$ in Figure \ref{fig:qin_venn}. Although there is moderate overlap among the selected feature sets, a substantial number of features are selected only by the method of \textcite{Dai2023FalseSplitting}.

\section{Discussion and conclusion}\label{s:disc}
In this paper, we introduced a novel FDR control framework based on data splitting. Our approach evaluates feature importance by comparing the predictive performance on the original data with that on the perturbed data.
The mirror statistic is symmetric about zero for null features, whereas relevant features tend to produce large positive values.
We estimate the number of false discoveries using the negative tail of the mirror-statistic distribution and, for each candidate threshold, estimate the FDR as the ratio of negative- to positive-tail counts exceeding that threshold. The final selected set corresponds to the smallest threshold that keeps the estimated FDR below the prespecified target.
Through extensive simulation studies, we demonstrated that the proposed method controls the FDR while maintaining high power compared with existing approaches for various situations.
When the signal–response relationship was linear, our method achieved power comparable to classical FDR procedures. Under nonlinear signal–response relationships, our approach successfully recovered true signals in scenarios where existing procedures failed to detect associations. Notably, in the most challenging high-dimensional non-linear setting, our method still achieved satisfying power. Overall, these results indicate that the proposed method provides a flexible and powerful framework for FDR control, with particular advantages in complex non-linear settings where classical approaches often struggle.

We also demonstrate the applicability of our method using a microbiome dataset. Existing approaches to handle compositionality often rely on linear models or log-ratio transformations. In contrast, our method predicts each feature conditional on the remaining features and then applies a closure-preserving step to maintain the compositional structure of the microbiome data. This allows us to select individual microbial features directly, rather than log-ratios relative to a reference taxon, thereby improving interpretability while accommodating complex feature--outcome relationships.
We acknowledge that modeling the conditional distribution of each feature using principal components derived from the remaining features may not fully capture complex dependence structures. However, this approach offers substantial computational efficiency.


A natural question is whether the proposed prediction-based, model-agnostic framework can be integrated with other feature selection approaches, such as the knockoff method. A practical challenge is that the standard knockoff procedure requires the original features and their corresponding knockoffs to be jointly included in the predictive model. Therefore, a dataset with $J$ original features requires fitting a model with $2J$ predictors. For machine learning models, especially in settings with limited sample size and high-dimensional microbiome features, this increase in dimensionality may further complicate model training.

Additionally, for Model-X knockoffs, an ideal implementation would use an independent external dataset to learn the covariate distribution and generate knockoff features, thereby separating knockoff construction from the data used for feature selection \cite{Barber2020RobustKnockoffs}; however, such external data are rarely available, and in practice the same dataset is often used both to estimate the covariate distribution for knockoff generation and to perform subsequent feature selection. In contrast, our data-splitting framework explicitly separates these roles, with 80\% of the observations used exclusively for model training and negative-control construction and the remaining 20\% used to evaluate the resulting feature statistics.

An alternative direction is to explore integration with e-value-based procedures, which do not require constructing knockoff variables. We would like to experiment with this integration in the future. Constructing valid e-values that satisfy the required null-expectation property remains an important methodological challenge.

\section{Acknowledgments}
This research was funded by the National Science Foundation, grant number NSF DMS 2310955.

\newpage

\newpage
\begin{table}[htbp]
\centering
\caption{Computation times in seconds for each method or component in a continuous-outcome simulation with \(n=1000\) and \(J=100\), measured on a MacBook Pro with an Apple M2 Max chip. The proposed MDS procedure used 10 random partitions and ten evaluation-fold rotations per partition, giving 50 outcome-prediction model fits.}
\label{tab:computation_time}
\begin{tabular}{p{3.2cm} p{9.5cm} r}
\toprule
\textbf{Method} & \textbf{Component} & \textbf{Seconds} \\
\midrule

Delporte MDS 
& Principal components construction 
& 6.14 \\

Delporte MDS 
& 50 XGBoost data-splitting fits and perturbation predictions 
& 13.86 \\

Dai MDS 
& 50 Lasso-screening + OLS data splits and MDS aggregation 
& 2.06 \\

Univariate BY 
& 100 univariate linear tests + BY adjustment 
& 0.007 \\

\midrule

Knockoff and Knockoff+ 
& Second-order knockoff construction 
& 0.072 \\

Knockoff and Knockoff+ 
& \texttt{glmnet} coefficient-difference statistic 
& 0.360 \\

Knockoff 
& Knockoff threshold (offset = 0) 
& $<0.001$ \\

Knockoff+ 
& Knockoff+ threshold (offset = 1) 
& 0.001 \\

\bottomrule
\end{tabular}
\end{table}

\newpage
\begin{table}[!htbp]
\centering
\caption{Microbial features selected by the proposed method in the analysis of the \textcite{Qin2014AlterationsCirrhosis} dataset. Because all features selected at $q=0.10$ were also selected at
$q=0.05$, features are categorized as those selected only at $q=0.05$ and those
selected at both FDR levels. Features selected by both the Lasso- and XGBoost-based selection procedures are marked in bold.}
\label{tab:qin_proposed_features}
\footnotesize
\begin{tabularx}{\textwidth}{XX}
\toprule
Selected only at $q=0.05$
&
Selected at both $q=0.05$ and $q=0.10$ \\
\midrule

\multicolumn{2}{l}{\textbf{Albumin: Lasso}} \\[0.3ex]

\textit{Bifidobacterium dentium};
\textit{Streptococcus parasanguinis}
&
\textit{Ruthenibacterium lactatiformans};
\textbf{\textit{Hungatella hathewayi}};
\textbf{\textit{Veillonella atypica}};
\textbf{\textit{Veillonella infantium}};
\textbf{\textit{Prevotella buccae}};
\textit{Prevotella} sp. CAG:520;
\textit{Ligilactobacillus salivarius}
\\[1.2ex]

\midrule
\multicolumn{2}{l}{\textbf{Albumin: XGBoost}} \\[0.3ex]

\textit{[Eubacterium] rectale};
\textit{Bacteroides xylanisolvens};
\textit{Prevotella stercorea};
\textit{Veillonella} sp. T11011-6;
\textbf{\textit{Prevotella buccae}}
&
\textit{Phocaeicola plebeius};
\textit{Prevotella copri};
\textbf{\textit{Hungatella hathewayi}};
\textit{Veillonella parvula};
\textit{Veillonella dispar};
\textbf{\textit{Veillonella atypica}};
\textbf{\textit{Veillonella infantium}};
\textit{Haemophilus} sp. HMSC71H05;
\textit{Dielma fastidiosa}
\\[1.2ex]

\midrule
\multicolumn{2}{l}{\textbf{Cirrhosis: Lasso}} \\[0.3ex]

\textbf{\textit{Haemophilus parainfluenzae}};
\textbf{\textit{[Clostridium] symbiosum}};
\textit{Streptococcus oralis};
\textit{Ligilactobacillus salivarius}
&
\textbf{\textit{Hungatella hathewayi}};
\textbf{\textit{Veillonella parvula}};
\textit{Adlercreutzia equolifaciens}; 
\textit{Veillonella atypica};
\textbf{\textit{Veillonella infantium}};
\textit{Phascolarctobacterium succinatutens};
\textit{Prevotella} sp. CAG:520;
\textit{Streptococcus} (genus)
\\[1.2ex]

\midrule
\multicolumn{2}{l}{\textbf{Cirrhosis: XGBoost}} \\[0.3ex]

\textit{Bacteroides uniformis};
\textit{Phascolarctobacterium faecium};
\textbf{\textit{[Clostridium] symbiosum}}
&
\textbf{\textit{Haemophilus parainfluenzae}};
\textbf{\textit{Hungatella hathewayi}};
\textbf{\textit{Veillonella parvula}};
\textbf{\textit{Veillonella infantium}}
\\

\bottomrule
\end{tabularx}
\end{table}
\newpage
\begin{table}[!htbp]
\centering
\caption{Number of microbial features selected in the analysis of the \textcite{Qin2014AlterationsCirrhosis} dataset. 
Results are shown for albumin level as a continuous outcome and cirrhosis status 
as a binary outcome at nominal FDR levels $q=0.05$ and $q=0.10$.}
\label{tab:qin_selection_number}
\begin{tabular}{lcccc}
\toprule
& \multicolumn{2}{c}{Albumin} 
& \multicolumn{2}{c}{Cirrhosis} \\
\cmidrule(lr){2-3}
\cmidrule(lr){4-5}
Method
& $q=0.05$ & $q=0.10$
& $q=0.05$ & $q=0.10$ \\
\midrule
Proposed (Lasso)   & 9  & 7  & 12 & 8 \\
Proposed (XGBoost) & 14 & 9  & 7  & 4 \\
Dai MDS            & 22 & 18 & 20 & 20 \\
Knockoff           & 0  & 0  & 3  & 3 \\
Knockoff+          & 0  & 0  & 0  & 0 \\
BY                 & 150 & 161 & 177 & 190 \\
\bottomrule
\end{tabular}
\end{table}
\newpage
\printbibliography[title={References}]
\end{refsection}
\begin{refsection}
\appendix

\section*{Supplementary Material}


\setcounter{page}{1}

\setcounter{figure}{0}

\renewcommand{\thefigure}{S.\arabic{figure}}
\section{Sensitivity analysis under feature correlation}

\label{sec:featureCorrelation}

\begin{figure}[!t]
    \centering
    \includegraphics[trim={0cm 0 0 1cm},clip,width=\linewidth]{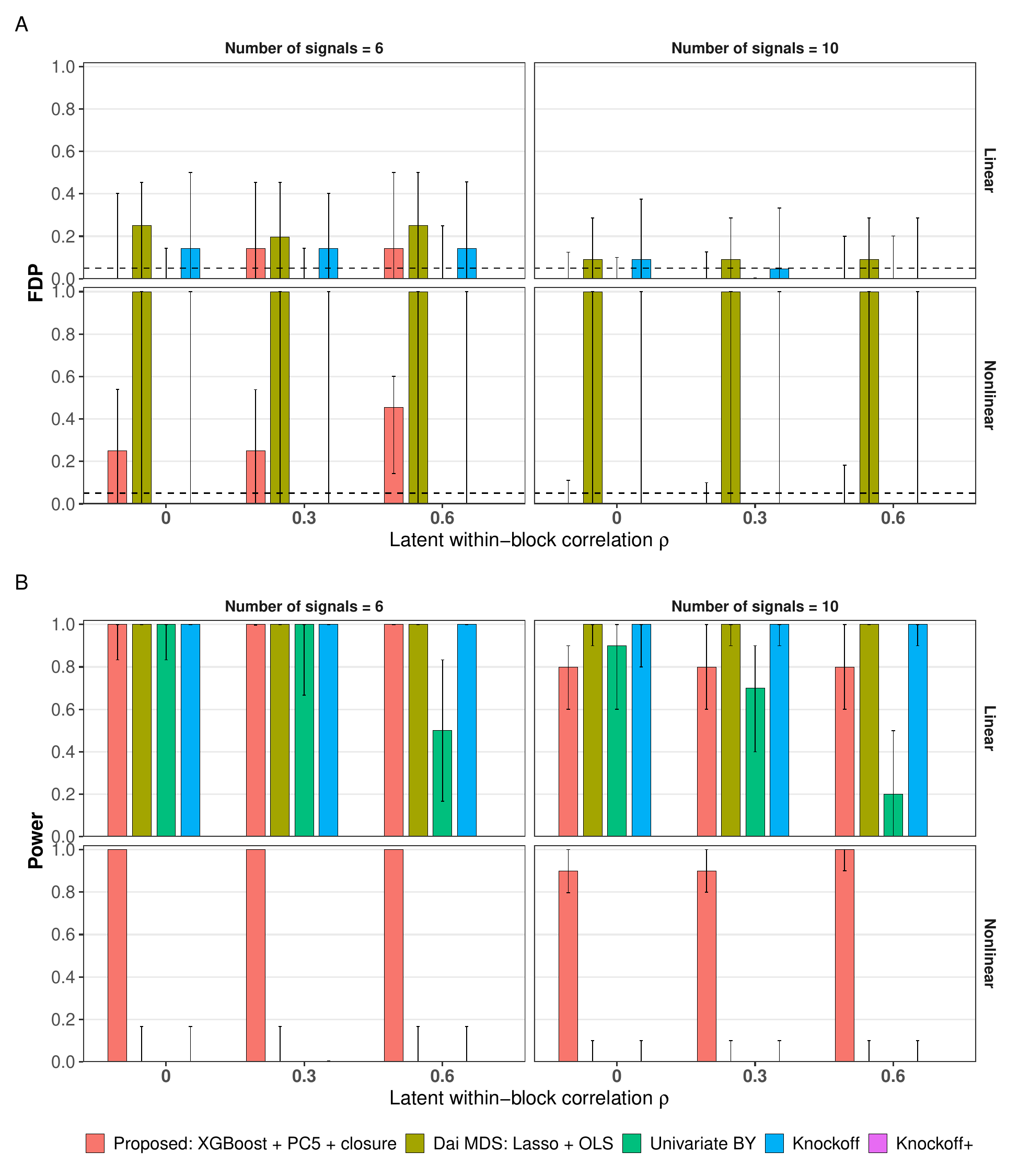}
    \caption{
        Median FDP (panel A) and median power (panel B) under a block correlation structure with Gaussian correlation $\rho \in \{0,0.3,0.6\}.$ Columns correspond to 6 and 10 true signals, and rows within each panel correspond to linear and nonlinear signal–outcome relationships. Bars show medians, and error bars show the $2.5$th and $97.5$th percentiles. Dashed horizontal lines in panel A indicate the corresponding nominal FDR level \(q=0.05\).
        The methods compared include the proposed MDS; the MDS procedure of \textcite{Dai2023FalseSplitting}; univariate BY; knockoff; and knockoff+.
        }
    \label{fig:featureCorrelation005}
\end{figure}

\begin{figure}[!t]
    \centering
    \includegraphics[trim={0cm 0 0 1cm},clip,width=\linewidth]{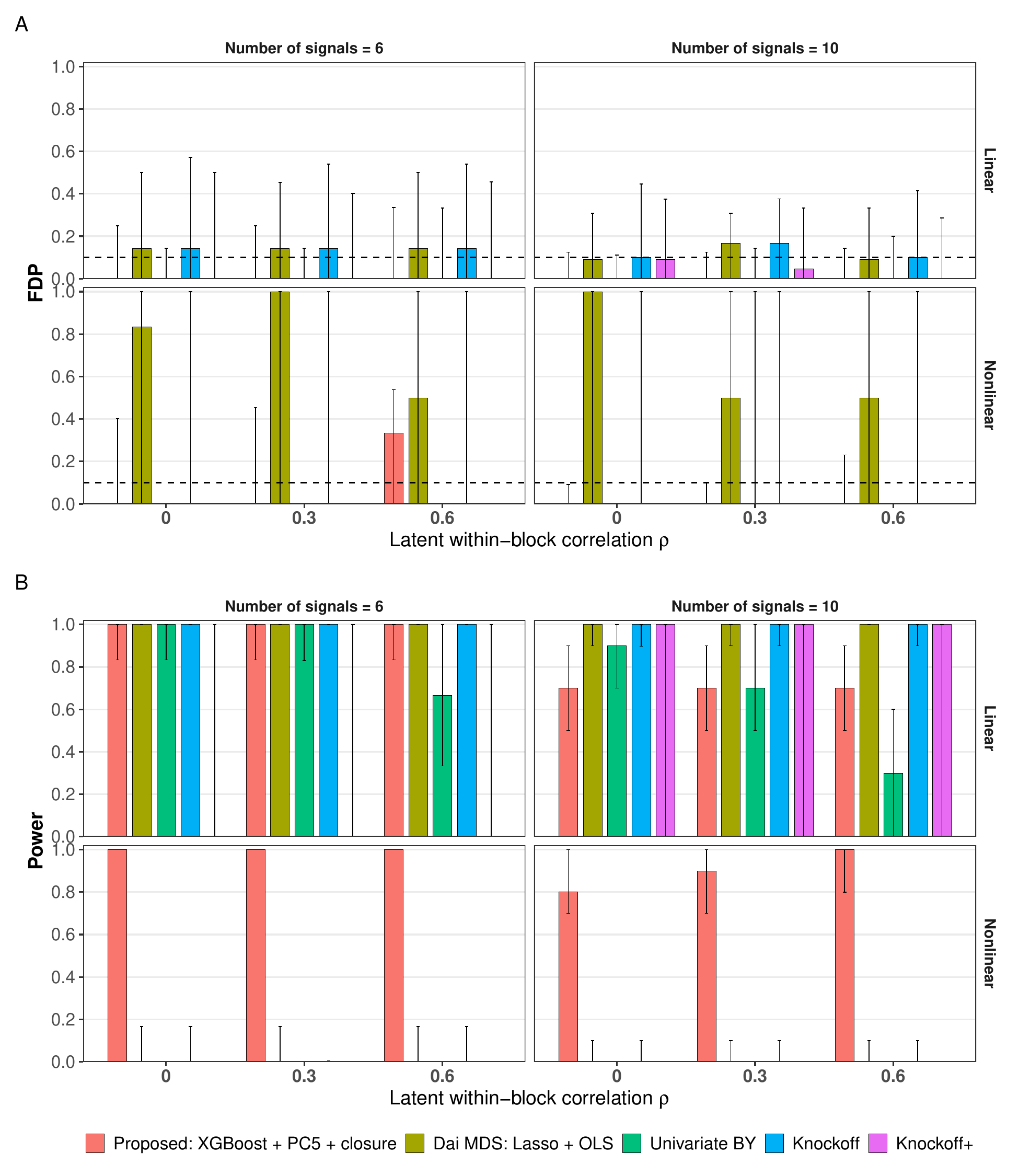}
    \caption{Median FDP (panel A) and median power (panel B) under a block correlation structure with Gaussian correlation $\rho \in \{0,0.3,0.6\}.$ Columns correspond to 6 and 10 true signals, and rows within each panel correspond to linear and nonlinear signal–outcome relationships. Bars show medians, and error bars show the $2.5$th and $97.5$th percentiles. Dashed horizontal lines in panel A indicate the corresponding nominal FDR level \(q=0.10\).
    The methods compared include the proposed MDS; the MDS procedure of \textcite{Dai2023FalseSplitting}; univariate BY; knockoff; and knockoff+.
    }
    \label{fig:featureCorrelation010}
\end{figure}

We conducted an additional sensitivity analysis to evaluate the performance of the proposed method under varying levels of positive feature correlation, with $J=100$ and $n=1000.$ Specifically, microbial compositions were generated from a block logistic-normal distribution. The latent Gaussian variables were divided into blocks of 10 features, with exchangeable within-block correlations set to $\rho \in \{0, 0.3, 0.6\}$ and zero correlation between blocks. The latent variables were then transformed to compositional relative abundances using the softmax transformation. We note that $\rho$ therefore represents the correlation among the latent Gaussian variables; because of the subsequent closure operation, the correlations among the observed relative abundances are not expected to be exactly equal to these values. 
Even when the latent within-block correlation is \(\rho=0\), the observed relative abundances can be negatively correlated after closure.
The outcome generation for linear and nonlinear settings follows the description for continuous outcomes in Section \ref{sec:simulationSetup}.
The performance of our method and the competing methods is shown in Figures \ref{fig:featureCorrelation005} and \ref{fig:featureCorrelation010}. The comparison of different methods for generating the negative controls under various correlation levels is presented in Section \ref{sec:compareNegativeControl}.

The correlation sensitivity analysis shows that the proposed method is generally robust to increasing feature dependence, particularly when the number of true features is considerable, and the target FDR is moderate. 
For linear relationships, \textcite{Dai2023FalseSplitting}'s MDS and the standard knockoff procedure achieve very high power, but this is accompanied by substantially larger empirical FDR in several settings. 
Knockoff+ provides stronger FDR protection, but is considerably more conservative: with six signals it has almost no power, and with ten signals it has essentially no power at $q=0.05$. Thus, under linear alternatives, the proposed method provides a favorable balance between the power and FDR control compared with existing methods.

One limitation is observed in the more sparse nonlinear setting with only six true signals. Although the proposed method maintains essentially 100\% power across all correlation levels, its empirical FDR increases as feature correlation becomes stronger.
This indicates that strong dependence can make finite-sample calibration of the mirror-statistic threshold more difficult when the number of true signals is small. 
Nevertheless, the competing methods provide little or no power in this nonlinear setting. Overall, the correlation sensitivity analysis indicates that the principal strength of the proposed method is its ability to preserve high power as feature correlation increases, particularly for nonlinear associations, while also showing that FDR calibration can become more challenging in very sparse nonlinear settings with strong feature dependence.

\section{Discussion on the finite-sample correction in FDR estimation}
\label{sec:numericalBuffer}

An interesting observation is that the number of selected features at a more stringent FDR threshold ($q=0.05$) can occasionally be larger than that obtained at a less stringent threshold ($q=0.10$) in the multiple data splitting framework. This phenomenon does not indicate a violation of FDR control, but rather results from the stochastic nature of the aggregation and thresholding procedure. 
In multiple data splitting methods, the final selection is not obtained from a single ordered list of test statistics. Instead, each feature is evaluated repeatedly across different data splits, producing a collection of split-specific evidence measures, which are subsequently aggregated to construct the final selection statistic. Because the aggregated statistics are random quantities affected by both the training/testing partition and the predictive model fitting, the ranking of features may vary slightly between different nominal FDR levels.
The estimated cutoff at $q=0.10$ is not necessarily identical to, or a relaxed version of, the cutoff at $q=0.05$. Random variation in the estimated null tail and the finite-sample resolution of the selection statistics can lead to situations where the $q=0.10$ cutoff is slightly more conservative than the $q=0.05$ cutoff for a particular dataset.

This phenomenon is expected to be more noticeable in high-dimensional settings with sparse signals. When only a small number of features have strong evidence of association, small changes in the estimated threshold can determine whether individual borderline features are included or excluded. Consequently, the number of discoveries may not be perfectly monotone with respect to the nominal FDR level in finite samples, even though the asymptotic FDR guarantee is maintained.

A practical consideration in implementing the mirror statistic framework is whether to include a finite-sample correction term in the estimated false discovery proportion (FDP). Specifically, one may estimate the FDP at threshold $t$ as
\[
\widehat{\mathrm{FDP}}(t)
=
\frac{1+\#\{j:M_j< -t\}}
{\#\{j:M_j> t\}\vee 1},
\]
where the additional one in the numerator provides a more conservative estimate, which is similar to the knockoff+ procedure \cite{CandFan18}. This correction is commonly motivated by finite-sample considerations and prevents overly optimistic FDP estimates when the number of selected features is small.



The finite-sample correction, although theoretically appealing, can substantially reduce power in sparse high-dimensional applications. Specifically, with the corrected FDP estimator,
any selected set satisfying $\widehat{\mathrm{FDP}}(t)\leq q$ must obey
\[
\#\{j:M_j>t\}\geq \frac{1}{q}.
\]
Any nonempty selected set must contain at least \(\lceil1/q\rceil\) features; otherwise the procedure returns no discoveries.
For example, at the commonly used FDR levels $q=0.05$ and $q=0.10$, the procedure cannot select fewer than 20 and 10 features, respectively. This requirement can be undesirable when the true signal is sparse, and the number of discoveries is expected to be small. 

In this work, we therefore use the asymptotic FDP estimator without the additional one in the numerator. Let $V(t)$ denote the number of false discoveries and $S(t)$ the total number of discoveries at threshold $t$. Under the null symmetry assumption,
\[
\mathbb{E}\{V(t)\}
=
\mathbb{E}\left[\#\{j:M_j\leq -t,H_j=0\}\right].
\]
Because the empirical proportion of null features converges as the number of hypotheses grows,
\[
\frac{\#\{j:M_j\leq -t\}}
{\#\{j:M_j\geq t\}}
\rightarrow
\frac{V(t)}{S(t)}
\]
in probability under standard weak dependence conditions. Therefore, the corresponding selection rule
\[
\widehat{S}
=
\{j:M_j\geq t_q\},
\]
where $t_q$ is the smallest threshold satisfying
\[
\frac{\#\{j:M_j\leq -t_q\}}
{\#\{j:M_j\geq t_q\}}
\leq q,
\]
preserves the target FDR asymptotically:
\[
\limsup_{J\rightarrow\infty}
\mathrm{FDR}
\leq q.
\]

\section{Comparison of methods for generating negative controls}
\label{sec:compareNegativeControl}

In most practical settings, users can hold the remaining \(J-1\) predictors fixed and construct a negative control for feature \(j\) by using those \(J-1\) predictors to model and generate the \(j\)th feature.
The validity of the prediction-based construction follows from the fact
that, under the conditional null, a negative control generated from the
correct conditional distribution of $X_j$ given $X_{-j}$ is
indistinguishable from the observed $X_j$ after conditioning on the
remaining covariates. Consequently, replacing the observed feature by
the generated negative control has no preferred direction of change in
predictive performance. When the conditional distribution is estimated
consistently, this exchangeability, and therefore the symmetry of the
null statistic, holds asymptotically.

We next provide a sufficient condition under which the prediction-based null and the required symmetry of the mirror statistic arise. Although our inferential target is the prediction-based null \(H_{0j}:P_{\mathrm{orig}}=P{(j)}\), Proposition~\ref{prop:prediction-based-control} considers the stronger conditional-independence condition \(Y\perp X_j\mid X_{-j}\). When the negative control is generated exactly from \(\mathcal L(X_j\mid X_{-j})\), conditional independence implies that the observed and generated versions of \(X_j\) are exchangeable conditional on \(X_{-j}\) and \(Y\). It therefore implies equality of the population predictive performances and, more importantly for the proposed procedure, symmetry of the corresponding prediction-performance difference. Conditional independence is thus sufficient, but not necessary, for the prediction-based null.

\begin{proposition}[Validity of a prediction-based negative control]
\label{prop:prediction-based-control}

For feature $j$, let
\(
F_j(\,\cdot\mid X_{-j})
=
\mathcal L(X_j\mid X_{-j})
\)
denote the true conditional distribution of $X_j$ given the remaining
covariates. Using data that are separate from the evaluation sample, fit
a prediction model
\(
\widehat F_j(\,\cdot\mid X_{-j})
\)
for this conditional distribution. For each observation in the evaluation
split $U_k$, generate
\(
\widetilde X_j^{(k)}
\sim
\widehat F_j(\,\cdot\mid X_{-j}^{(k)})
\)
independently of $X_j^{(k)}$ and $Y^{(k)}$ conditional on
$X_{-j}^{(k)}$.
Suppose that feature $j$ is conditionally null, so that
\(
Y\perp X_j\mid X_{-j}.
\)
If the prediction model correctly captures the conditional distribution,
that is,
\(
\widehat F_j(\,\cdot\mid X_{-j})
=
F_j(\,\cdot\mid X_{-j}),
\)
then the split-specific performance difference
\(
\beta_j^{(k)}
=
\mathcal P^{(k)}-\mathcal P_j^{(k)}
\)
is symmetric about zero.
\end{proposition}

\begin{proof}
Under the conditional-null hypothesis,
\(
Y\perp X_j\mid X_{-j},
\)
the conditional distribution of $X_j$ is unchanged after additionally
conditioning on the outcome. Therefore,
\(
\mathcal L(X_j\mid X_{-j},Y)
=
\mathcal L(X_j\mid X_{-j})
=
F_j(\,\cdot\mid X_{-j}).
\)
Because the prediction model correctly captures this conditional
distribution, the generated negative control satisfies
\(
\mathcal L(\widetilde X_j\mid X_{-j},Y)
=
F_j(\,\cdot\mid X_{-j}).
\)
Moreover, $\widetilde X_j$ is generated independently of $X_j$ and $Y$
conditional on $X_{-j}$. Hence, conditional on $X_{-j}$ and $Y$,
$X_j$ and $\widetilde X_j$ are independent draws from the same
distribution. It follows that
\(
(X_j,\widetilde X_j)\mid X_{-j},Y
\stackrel{d}{=}
(\widetilde X_j,X_j)\mid X_{-j},Y.
\)
Now treat the fitted outcome-prediction model as fixed when evaluating
split $U_k$. Let
$\mathcal P^{(k)}$
and $\mathcal P_j^{(k)}$ denote the original predictive performance and the predictive performance obtained
when the $j$th covariate in the evaluation sample is set equal to $\widetilde X_j$,
while the remaining covariates and the outcome are unchanged, respectively. 
Because $X_j^{(k)}$ and $\widetilde X_j^{(k)}$ are exchangeable,
\(
\left(
\mathcal P^{(k)},
\mathcal P_j^{(k)}
\right)
\stackrel{d}{=}
\left(
\mathcal P_j^{(k)},
\mathcal P^{(k)}
\right).
\)

Consequently, for every $\delta>0$,
\begin{align*}
\mathbb P\left(\beta_j^{(k)}>\delta\right)
&=
\mathbb P\left(
\mathcal P^{(k)}-\mathcal P_j^{(k)}>\delta
\right)\\
&=
\mathbb P\left(
\mathcal P_j^{(k)}-\mathcal P^{(k)}>\delta
\right)\\
&=
\mathbb P\left(
\mathcal P^{(k)}-\mathcal P_j^{(k)}<-\delta
\right)\\
&=
\mathbb P\left(\beta_j^{(k)}<-\delta\right).
\end{align*}
Therefore,
\(
\beta_j^{(k)}
\stackrel{d}{=}
-\beta_j^{(k)},
\)
and $\beta_j^{(k)}$ is symmetric about zero.
Conditional on the training data, the two evaluation sets are independent, and \(\beta_j^{(1)}\) and \(\beta_j^{(2)}\) are independently symmetric about zero. Therefore,
\((\beta_j^{(1)},\beta_j^{(2)})\overset d=(-\beta_j^{(1)},\beta_j^{(2)})\).
Since \(M_j(-a,b)=-M_j(a,b)\), the conditional distribution of \(M_j\) is symmetric.
\end{proof}

\begin{figure}[!h]
    \centering
    \includegraphics[width=\linewidth]{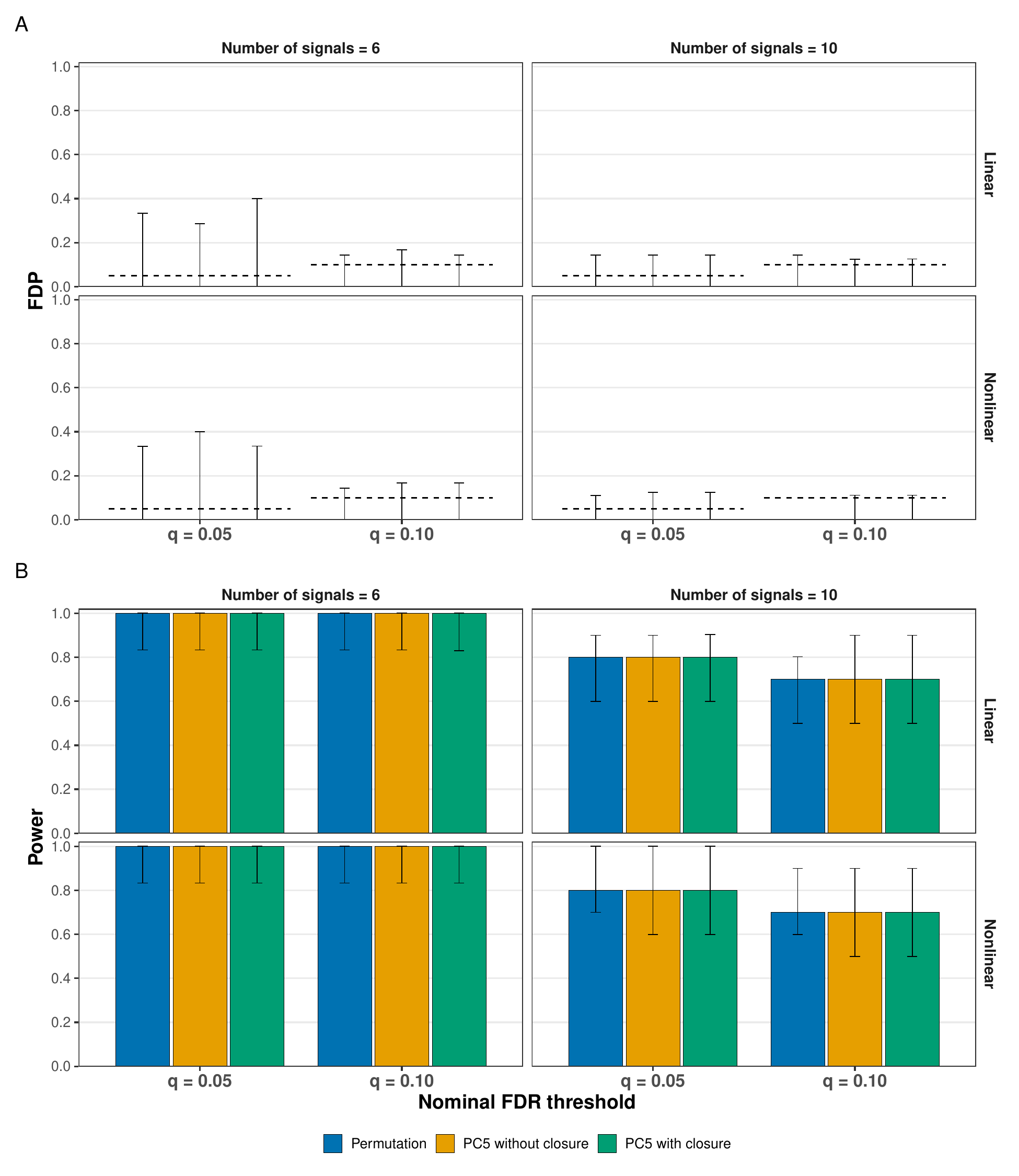}
    \caption{Median FDP (panel A) and median power (panel B) with features generated from a Dirichlet distribution.  Bars show medians, and error bars show the $2.5$th and $97.5$th percentiles. Dashed horizontal lines in panel A indicate the corresponding nominal FDR level \(q\).
    The negative control generation methods compared include permutation, prediction based on the first 5 principal components, and prediction based on the first 5 principal components and compositionality closure adjustment.}
    \label{fig:perturbationMethodComparisonDirichlet}
\end{figure}

\begin{figure}[!h]
    \centering
    \includegraphics[width=\linewidth]{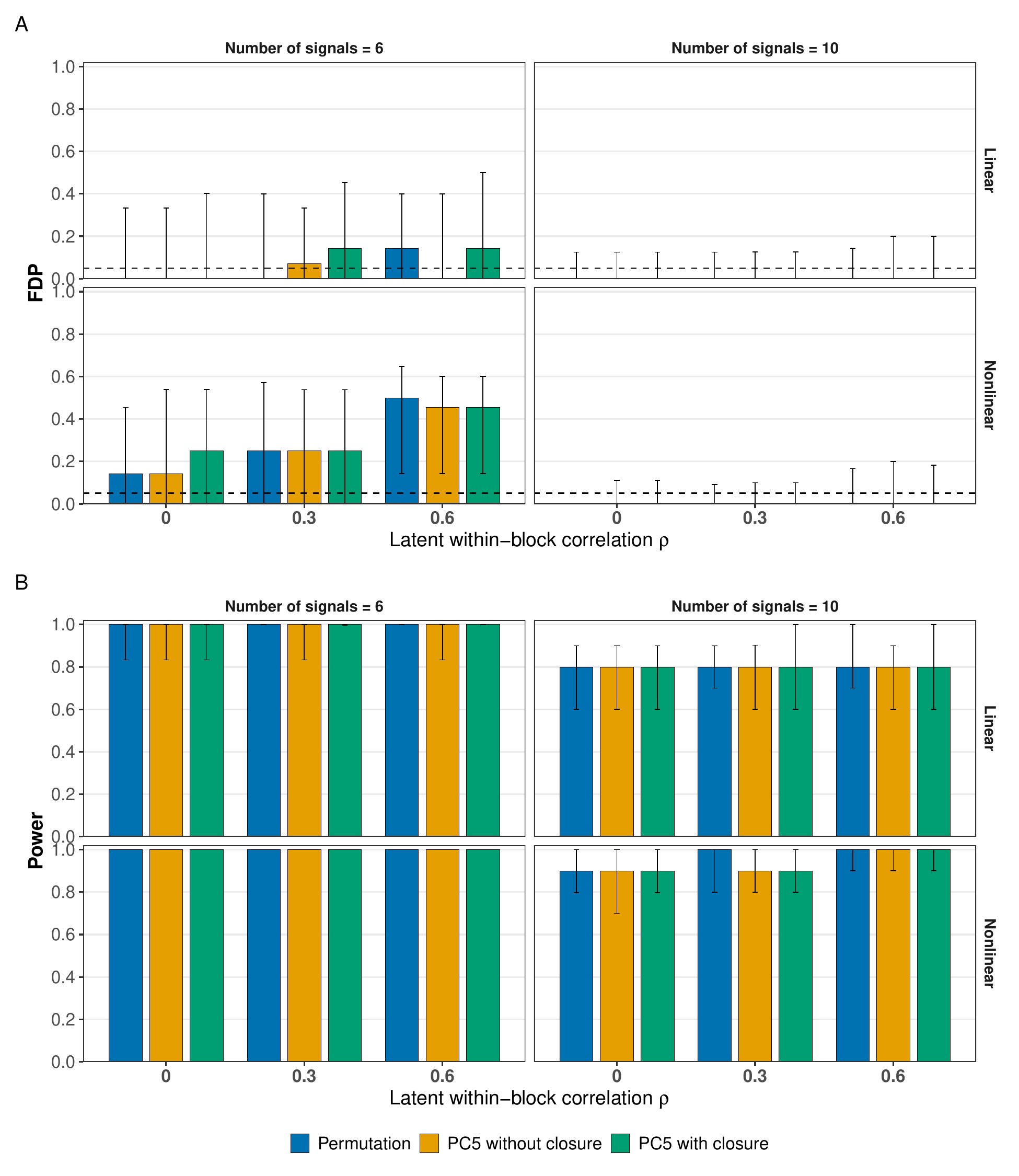}
    \caption{
    Median FDP (panel A) and median power (panel B) with features generated from latent Gaussian blocks with $\rho \in \{0,0.3,0.6\}$.  Bars show medians, and error bars show the $2.5$th and $97.5$th percentiles. Dashed horizontal lines in panel A indicate the corresponding nominal FDR level \(q=0.05\).
    The negative control generation methods compared include permutation, prediction based on the first 5 principal components, and prediction based on the first 5 principal components and compositionality closure adjustment.
    }
    \label{fig:perturbationMethodComparisonLatentGaussian005}
\end{figure}

\begin{figure}[!h]
    \centering
    \includegraphics[width=\linewidth]{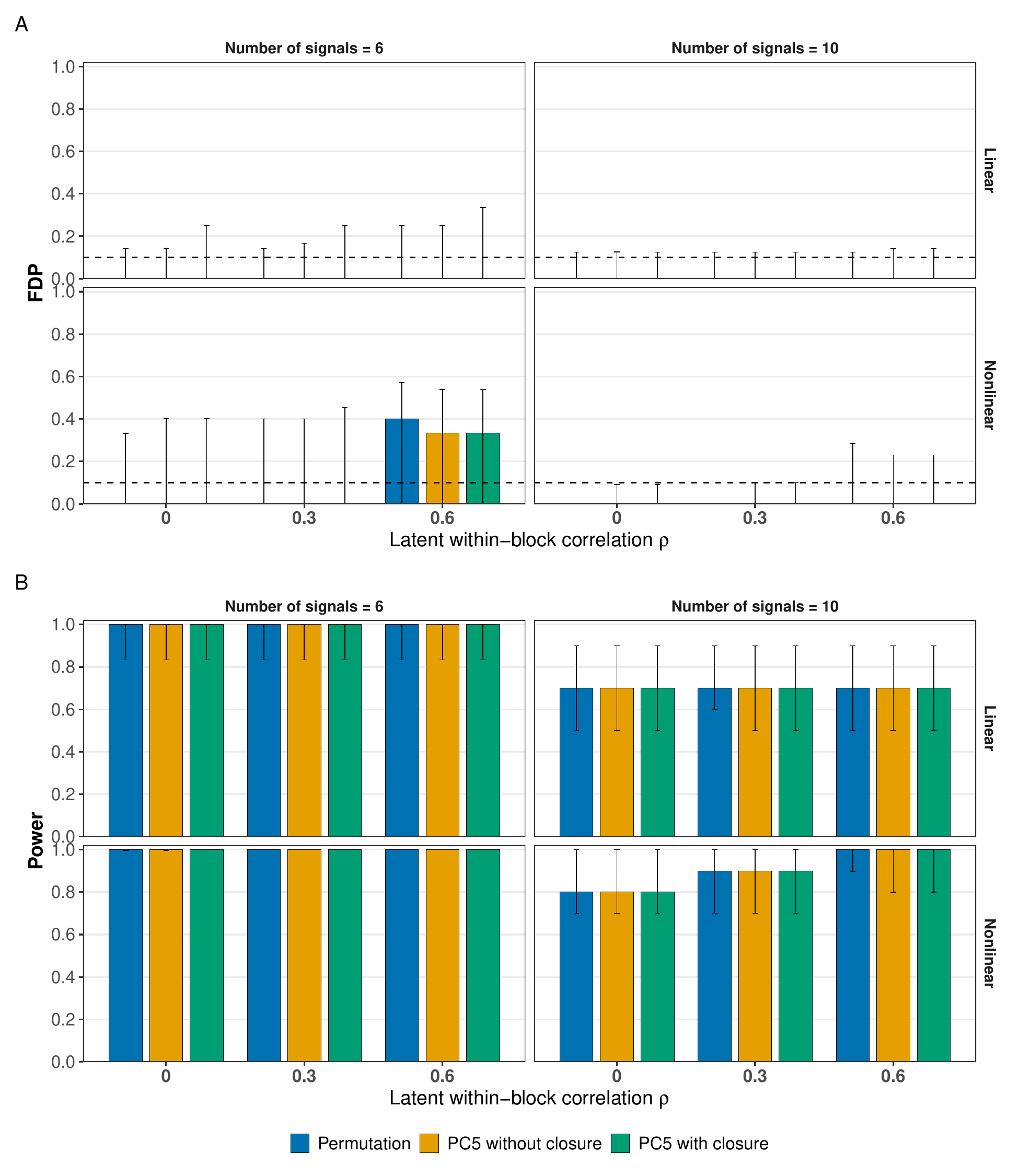}
    \caption{    Median FDP (panel A) and median power (panel B) with features generated from latent Gaussian blocks with $\rho \in \{0,0.3,0.6\}$.  Bars show medians, and error bars show the $2.5$th and $97.5$th percentiles. Dashed horizontal lines in panel A indicate the corresponding nominal FDR level \(q=0.10\).
    The negative control generation methods compared include permutation, prediction based on the first 5 principal components, and prediction based on the first 5 principal components and compositionality closure adjustment.}
    \label{fig:perturbationMethodComparisonLatentGaussian010}
\end{figure}

Proposition~\ref{prop:prediction-based-control} establishes this relationship for the exact conditional-generation construction in which \(X_{-j}\) is held fixed. For the closure-preserving compositional construction described in Section \ref{sec:constructNegative}, the perturbation also modifies the remaining components, and the proposition therefore does not apply exactly.
However, in high-dimensional microbiome data, the adjustment required to preserve the compositional constraint is distributed across the other \(J-1\) features. Consequently, the induced change in the abundance of each remaining feature is typically small, so the overall perturbation to the rest of the feature vector is expected to be limited.
We investigate the resulting approximation empirically in the following simulation study.
We compared three negative-control constructions:
(i) marginal permutation, whose validity requires $X_j\perp (Y,X_{-j})$; (ii) conditional generation using beta regression,
with the first five principal components of the remaining features as
predictors; and (iii) the same principal-component-based conditional
generation followed by closure adjustment.
As shown in Figures~\ref{fig:perturbationMethodComparisonDirichlet}, \ref{fig:perturbationMethodComparisonLatentGaussian005}, and \ref{fig:perturbationMethodComparisonLatentGaussian010}, the operating characteristics
were highly similar across the three approaches.
In particular, enforcing closure did not materially improve the empirical
performance of the proposed procedure in the simulation settings considered.

In our simulation and real data application, we used the principal-component-based conditional
construction with closure adjustment as the default implementation for microbiome data. 
The closure-unadjusted construction is also provided in our GitHub for
non-compositional high-dimensional data.

\printbibliography[title={Supplemental References}]

\end{refsection}

\end{document}